\documentclass[a4paper,fleqn]{cas-sc}

\usepackage[numbers]{natbib}
\usepackage{amsmath,amsfonts,amssymb}
\usepackage{algorithm}
\usepackage{mathtools}
\usepackage{array}
\usepackage{algpseudocode}
\usepackage{textcomp}
\usepackage{stfloats}
\usepackage{url}
\usepackage{verbatim}
\usepackage{graphicx}
\usepackage{subcaption}
\def\tsc#1{\csdef{#1}{\textsc{\lowercase{#1}}\xspace}}
\tsc{WGM}
\tsc{QE}
\tsc{EP}
\tsc{PMS}
\tsc{BEC}
\tsc{DE}
\begin{document}
\let\WriteBookmarks\relax
\def\floatpagepagefraction{1}
\def\textpagefraction{.001}

\shorttitle{iCNN-LSTM: A batch-based incremental ransomware detection system using Sysmon}

\shortauthors{Ispahany et~al.}

\title [mode = title]{iCNN-LSTM: A batch-based incremental ransomware detection system using Sysmon}                      

%\tnotetext[1]{This document is the results of the research
%   project funded by the National Science Foundation.}

%\tnotetext[2]{The second title footnote which is a longer text matter
%   to fill through the whole text width and overflow into
%   another line in the footnotes area of the first page.}

\author[1,3]{Jamil Ispahany}[type=editor,
                        auid=000,bioid=1,
                        orcid=0000-0001-8224-2924]

%\cormark[1]


\ead{Jispahany@csu.edu.au}

%\ead[url]{www.cvr.cc, cvr@sayahna.org}

%\credit{Conceptualization of this study, Methodology, Software}

\affiliation[1]{organization={Charles Sturt University}, 
    city={Bathurst},
    % citysep={}, % Uncomment if no comma needed between city and postcode 
    state={NSW},
    postcode={2795},
    country={Australia}}

\author[2,3]{MD Rafiqul Islam}

\author[1,3]{M. Arif Khan}
%\ead{cvr3@sayahna.org}
%\ead[URL]{www.sayahna.org}

%\credit{Data curation, Writing - Original draft preparation}

\affiliation[2]{organization={Charles Sturt University}, 
    city={Albury-Wodonga},
    % citysep={}, % Uncomment if no comma needed between city and postcode
    state={NSW},
    postcode={2640},
    country={Australia}} 

\author%
[1,3]
{MD Zahidul Islam}
%\cormark[2]
%\fnmark[1,3]
%\ead{rishi@stmdocs.in}
%\ead[URL]{www.stmdocs.in}

\affiliation[3]{organization={Cyber Security Cooperative Research Centre}, 
    city={Kingston},
    % citysep={}, % Uncomment if no comma needed between city and postcode 
    state={ACT},
    postcode={2600},
    country={Australia}}

\cortext[cor1]{Corresponding author}
%\cortext[cor2]{Principal corresponding author}

%\fntext[fn1]{This is the first author footnote. but is common to third
%  author as well.}
%\fntext[fn2]{Another author footnote, this is a very long footnote and
%  it should be a really long footnote. But this footnote is not yet
%  sufficiently long enough to make two lines of footnote text.}

%\nonumnote{This note has no numbers. In this work we demonstrate $a_b$
%  the formation Y\_1 of a new type of polariton on the interface
%  between a cuprous oxide slab and a polystyrene micro-sphere placed
%  on the slab.
%  }

\begin{abstract}
In response to the increasing ransomware threat, this study presents a novel detection system that integrates Convolutional Neural Networks (CNNs) and Long Short-Term Memory (LSTM) networks. By leveraging Sysmon logs, the system enables real-time analysis on Windows-based endpoints. Our approach overcomes the limitations of traditional models by employing batch-based incremental learning, allowing the system to continuously adapt to new ransomware variants without requiring complete retraining. The proposed model achieved an impressive average F2-score of 99.61\%, with low false positive and false negative rates of 0.17\% and 4.69\%, respectively, within a highly imbalanced dataset. This demonstrates exceptional accuracy in identifying malicious behaviour. The dynamic detection capabilities of Sysmon enhance the model's effectiveness by providing a reliable stream of security events, mitigating the vulnerabilities associated with static detection methods. Furthermore, the parallel processing of LSTM modules, combined with attention mechanisms, significantly improves training efficiency and reduces latency, making our system well-suited for real-world applications. These findings underscore the potential of our CNN-LSTM framework as a robust solution for real-time ransomware detection, ensuring adaptability and resilience in the face of evolving cyber threats.
\end{abstract}

% \includegraphics{figs/grabs.pdf}

%\input{highlights}

% Keywords

%\------------------------------------------------------------------------------------
\maketitle
\section{Introduction}

Since the onset of the COVID-19 crisis, malware has emerged as a significant global challenge \cite{ispahany2021detecting}. Among various types of malware, ransomware has recently become more infamous. Ransomware is malicious software that encrypts the victim's data, rendering it inaccessible until a ransom is paid, typically demanded in cryptocurrency. This form of cyberattack has become prolific due to its lucrative nature for perpetrators and the relative ease of deployment compared to other cybercrimes. These attacks lead to financial losses, disrupt essential services, and compromise sensitive data. The severity of ransomware attacks was underscored on May 7, 2021, when Colonial Pipeline Co., the largest oil pipeline operator in the United States, fell victim to an attack by the Darkside hacking group. The hackers infiltrated the network, stole over 100 gigabytes of data, and then encrypted the company's network, demanding over USD 4.4 million ransom \cite{beerman2023review}. In response to the attack, Colonial Pipeline temporarily halted its operations. This disruption led to significant supply shortages and logistical challenges across the East Coast, contributing to a sharp increase in gas prices \cite{dossett_2021}.\\
The research community has focused on developing more effective detection methods in response to the escalating ransomware threat. Historically, early research primarily utilised signature-based approaches that depended on known malware signatures for detection. While straightforward, these methods often missed new or altered ransomware strains. To overcome these limitations, behaviour-based approaches have become more favoured, as they analyse system activity patterns to identify malicious behaviours and are more adept at detecting zero-day attacks.\\
Machine learning techniques have been instrumental in automating and refining ransomware detection strategies \cite{ispahany2024ransomware}. Initially, research concentrated on traditional machine learning methods because they were straightforward to implement, used resources efficiently, and yielded dependable outcomes. However, in recent times, there has been a shift towards deep learning techniques in ransomware detection. This change is attributed to deep learning's enhanced performance \cite{bello2021detecting}, proficiency in identifying complex patterns, capability to handle large datasets \cite{fernando2020study}, and ability to generalise effectively to new, unseen data \cite{hemalatha2021efficient}, thereby instilling confidence in their effectiveness.\\
Deep learning methods have shown promising results in detecting ransomware behaviour, leveraging their strength in deriving insights from complex and sequential data types. Despite these advancements, a significant challenge remains: most models in the literature are built from scratch, which is not only resource-intensive but also inefficient, given the rapid proliferation of ransomware variants \cite{ispahany2024ransomware}. Incremental model updating offers a solution to this issue by allowing existing models to adapt to new data without retraining the model from scratch, thereby effectively maintaining knowledge of new and historical ransomware threats.\\
\begin{comment}
\begin{itemize}
    \item \textit{Ransomware Detection from Sysmon Logs in real-time :} We demonstrate how deep learning techniques can detect ransomware activity from a continuous live stream of Sysmon logs from Windows-based endpoints, leveraging spatial and temporal data patterns.
    \item \textit{Novel Incremental Detection System:} We propose an incremental ransomware detection system called iCNN-LSTM that combines convolutional neural networks (CNNs) with long short-term memory (LSTM) networks. Our system incrementally updates its model in batches, eliminating the need for retraining from scratch while maintaining high detection accuracy.
    \item \textit{Enhanced Model Efficiency:} We evaluated other non-incremental CNN and LSTM models from the literature by running them through our batch-incremental framework and measuring their classification performance on our dataset for benchmarking. Our results demonstrate that our parallel CNN-LSTM architecture improves detection accuracy, reduces false negatives, and enhances runtime performance, even in a highly imbalanced dataset where ransomware is the minority class.
\end{itemize}
\end{comment}
In response to these challenges, we offer the following contributions:
\begin{itemize}
    \item \textit{Novel detection architecture:} We present a system for efficient ransomware detection using Sysmon log streams. Our method achieves an impressive F2-score of 99.61\% and a low false positive rate of 4.69\% in an imbalanced dataset where ransomware is the minority class, demonstrating superior performance in detecting ransomware threats.
    \item \textit{Continuous Learning Mechanism:} We implement a continuous learning approach that updates the model with mini-batches of data, improving its adaptability to new ransomware strains and addressing class imbalance through SMOTE. Compared to other incremental learning-based ransomware detection systems in the literature, our technique achieves the highest F1 and F2 scores.
    \item \textit{Efficient Processing:} We propose a CNN-LSTM deep-learning architecture that leverages parallel LSTM configurations and attention mechanisms to reduce runtime and enhance processing efficiency. We compare our approach with other CNN-LSTM architectures from the literature using our incremental framework. We show that it achieves the highest F2-score, recall, and precision, along with the lowest false negative rate and fastest runtime relative to the F2-score. This makes it highly suitable for real-time applications.
\end{itemize}

Our model is designed for adaptability, incorporating incremental updates that enable it to evolve continuously in response to emerging threats. This dynamic updating mechanism allows the system to stay updated without completely rebuilding the model from scratch. It offers accurate and timely classification, ensuring it remains effective against the latest ransomware tactics.\\

%\------------------------------------------------------------------------------------
\section{Related work}\label{sec:related-work} 

\subsection{Ransomware detection using CNNs and LSTMs}

Numerous studies have employed deep learning methods to identify ransomware activities such as LSTMs \cite{karbab2023swiftr}, CNNs \cite{zhang2020ransomware}, GAN \cite{zhang2021dual, gazzan2023enhanced} and Autoencoders \cite{zahoora2022ransomware}. Long Short-Term Memory Networks (LSTMs), developed by Hochreiter and Schmidhuber, remain a popular choice amongst ransomware detection studies and are particularly effective in recognising patterns in sequential data, such as those found in runtime logs. These networks were designed to overcome the limitations of traditional Recurrent Neural Networks (RNNs), such as the exploding gradient problem during backpropagation through time. LSTMs are adept at preserving information over long periods due to their unique structure, which includes three gates: input, forget, and output. These gates respectively manage the influx of new data, the retention or removal of existing data, and the output calculations based on the current state of the cell.\\
LSTMs have been widely utilised in ransomware detection. For instance, Woralert et al. \cite{woralert2023hard} developed a real-time detection framework for hardware anomalies indicative of ransomware. The authors use an LSTM-based predictor combined with ensemble methods and moving averages to capture the temporal aspects of hardware-level data from performance monitoring counters essential for identifying ransomware attacks.\\
Roy and Chen \cite{roy2021deepran} employ BiLSTMs to detect anomalies related to ransomware activity. While the model achieves perfect precision, their system is trained exclusively on goodware, leading to a bias towards the majority class. This is reflected in its precision and recall rates, which indicate many false negatives for ransomware detection. The model's performance suggests an inability to handle class imbalance effectively. \\
Convolutional Neural Networks (CNNs) are another deep learning technique frequently used in image and video recognition and adapted for ransomware detection. CNNs excel at learning hierarchical features from input data and identifying structural patterns in malware samples. For example, Ciaramella et al. \cite{ciaramella2023explainable} achieved notable success using CNNs to distinguish between malicious and benign software by analysing visual data derived from binary code, reaching a high accuracy rate. However, this method primarily relies on static features from the binary code, which may limit its effectiveness in dynamic scenarios.\\
Numerous studies have realised the benefits of combining deep learning algorithms like Convolutional Neural Networks (CNNs) and Long Short-Term Memory (LSTMs) networks within the ransomware and malware detection domains \cite{}. This hybrid architecture enhances performance by leveraging the strengths of both CNNs and LSTMs \cite{luan2019research}. Specifically, CNNs are adept at extracting spatial features, while LSTMs excel at recognising temporal patterns in sequential data. These characteristics make the combined CNN-LSTM architecture particularly effective for detecting ransomware activities in time-series logs.\\
Bensaoud and Kalita present a novel approach to malware classification by integrating Convolutional Neural Networks (CNN) and Long Short-Term Memory (LSTM) networks, achieving a remarkable accuracy of 99.91\%. The authors employ advanced feature extraction techniques, including Bag-of-Words (BoW), Term Frequency-Inverse Document Frequency (TF-IDF), and One-hot Encoding, to effectively represent API calls and opcode sequences. A significant strength of the study lies in its empirical evaluation against various pre-trained models, showcasing the efficacy of the proposed CNN-LSTM architecture. However, this article has limitations, including the reliance on a highly imbalanced dataset, where ransomware samples make up the majority class. Furthermore, the proposal does not provide performance metrics such as runtime and cannot detect ransomware in real-time data streams, which limits its applicability in real-world scenarios.\\
Akhtar and Feng also present a novel approach to malware detection: integrating single-layer Convolutional Neural Networks (CNN) and Long-Short-Term Memory (LSTM) networks. The study demonstrates significant improvements in detection accuracy, achieving a reported accuracy of 99.6\% with a low false positive rate, indicating the model's robustness in identifying malicious software. However, much like the previous study, the research is limited by its reliance on a small, imbalanced dataset, which may only partially represent the diversity of real-world malware. Additionally, there is no commentary on the runtime of the model.\\
These studies demonstrate the enhanced classification accuracy achieved by combining CNNs and LSTMs to analyse spatial and temporal features. However, in real-world applications, ransomware is typically a minority class. These studies do not address the challenge of detecting ransomware within an imbalanced, continuous data stream where ransomware is underrepresented. Rapid ransomware classification is crucial for protecting sensitive data and has not been explored in previous studies. Finally, no other studies have compared CNN-LSTM architectures, highlighting a unique opportunity for further investigation. With this context in mind, our study addresses these key objectives.

\subsection{Ransomware Detection Systems using incremental learning techniques}

Many machine-learning-based ransomware detection proposals throughout the literature require the underlying model to be rebuilt from scratch to update it with new knowledge \cite{ispahany2024ransomware}. This approach is increasingly impractical as the frequency of new ransomware variants on the internet continues to rise, necessitating frequent retraining of models. Such retraining is time-consuming and resource-intensive, given the significant computational demands of deep learning and its lengthy training processes \cite{yang2022adaptability}. However, updating the underlying model frequently is critical, particularly when protecting sensitive data from new ransomware strains. An underexplored alternative in ransomware detection is incremental learning to update the underlying model. Incremental learning approaches have been successfully deployed in other domains, such as the energy sector \cite{shaohu2024prediction}. However, their adoption in ransomware detection has been limited. \\
Updating the underlying model frequently can keep the detection system current without complete retraining, thus preserving resources and improving responsiveness. There are two general approaches to updating the model from data streams incrementally: instance-incremental and batch-based incremental learning. Instance-based incremental learning approaches require the model to be trained on each inbound data instance, making them susceptible to data distribution changes over time (otherwise known as concept drift) \cite{read2012batch} and forgetting old information when updating the model with new information (otherwise known as catastrophic forgetting) \cite{gepperth2016incremental}. On the other hand, batch incremental learning builds models from data batches and removes older batches once memory capacity is reached. This allows the model to phase out outdated data while learning from new data, offering adaptability to concept drift within the dataset \cite{read2012batch}
Within data streams, catastrophic forgetting can result in poor classification of the minority class within incremental learning models if not properly addressed. New classes replace old ones, and only a limited number of older samples can be retrained, leading to class imbalance. Class imbalance can bias predictions toward the majority class, leading to poor recognition of minority classes and often resulting in a low recall score. \cite{wallace2011class, belouadah2020active}. This is particularly problematic in ransomware detection, where failure to detect the minority class (ransomware events) can have severe consequences. \\ 
This phenomenon was observed in the study conducted by Roy and Chen \cite{roy2021deepran}, who introduced a ransomware detection system built on a Bidirectional LSTM and Conditional Random Field (CRF) model to identify 17 types of ransomware attacks. In their study, the authors used online learning techniques to detect real-time anomalies from ransomware activity within a data stream. However, whilst the model achieves perfect precision, their system is trained exclusively on goodware, leading to class imbalance. This is reflected in its recall rates, which indicate a disproportionate number of false negatives for the minority class since only 7 out of 17 detected ransomware types had a recall rate above 90\% for most ransomware types. The model's performance suggests an inability to handle class imbalance effectively. \\
Al-rimy et al. \cite{al2019crypto} adopt the alternative approach and develop a batch-based incremental bagging technique utilising non-deep learning methods. Their study divided the dataset into "sub-spaces" to extract detailed features from temporal instances within batches. The authors extract more contextual information from the data by updating their model in batches, such as TF-IDF. They employed an ensemble of algorithms to classify the dataset, achieving an F1-score of 98.7\%. Despite the high accuracy, the system is not designed to detect ransomware within a continuous stream of logs, which limits its real-time detection capabilities.\\
Batch-based incremental learning provides a balanced approach, leveraging the enhanced temporal statistics from batch learning while reducing the concept drift and catastrophic challenges typically associated with instance-based incremental learning within data streams. This is particularly advantageous for facilitating real-time detection. However, despite its utility, its use amongst data streams within the ransomware detection domain has been limited. This suggests a potential area for further research and application.

%\------------------------------------------------------------------------------------

\section{Proposed methodology}\label{sec:approach}

%\subsection{Overview of our proposed architecture}
\subsection{Overview of our proposed framework}

%\----------------------------------------------------------
% Figure: Proposed architecture
%\----------------------------------------------------------
\begin{figure}[ht!]
    \centering
    \includegraphics[width=1.00\linewidth]{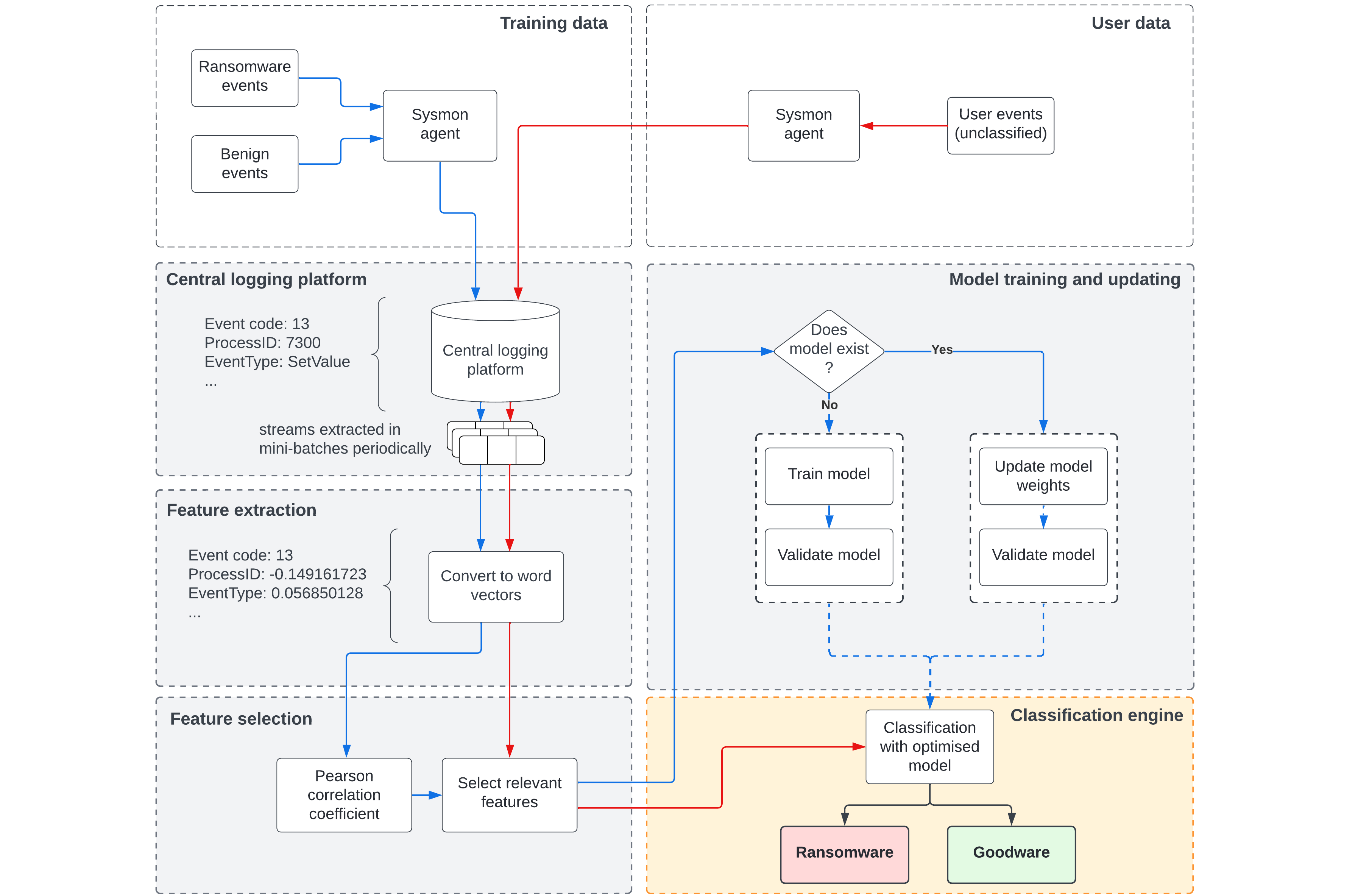}
    \caption{The proposed batch-based incremental ransomware detection framework using Sysmon. Blue arrows indicate the data stream that trains the model, while red arrows represent the data flow to be classified. Additionally, the blue dashed line symbolises the logical step of baselining the model, distinguishing it from actual data flows. }
    \label{fig:architecture}
\end{figure}
%\----------------------------------------------------------

Fig. \ref{fig:architecture} provides an overview of the operational pipeline of our ransomware detection system, which leverages a convolutional neural network (CNN) combined with a long short-term memory (LSTM) model for classification. The process begins with Sysmon agents, a monitoring tool that captures system events related to security breaches and normal operations. These events are then transmitted to a centralised logging platform that aggregates the information and organises it into mini-batches. These mini-batches are periodically dispatched to the ransomware detection system, ensuring it is frequently updated with the most current data. This is critical for maintaining its relevance and accuracy in the fast-evolving landscape of cybersecurity threats. \\
Events are passed through the feature extraction stage, where events are converted into word vectors, laying the groundwork for effective classification. Feature selection is refined using the Pearson correlation coefficient to pinpoint features significantly impacting the classification results.\\
The selected features are directed to the classification model, designed to adapt dynamically to new patterns and anomalies in the data. The initial model is built using the data processing techniques discussed in section \ref{dataset-preparation} and is constantly updated once deployed. It achieves this through periodic weight updates, incorporating the latest insights from the newly processed mini-batches. This continuous learning approach allows the model to evolve in response to the emergence of new ransomware signatures or variations in goodware behaviour, maintaining its effectiveness in real-time threat detection and classification.

% --------------------------------------
% SYSMON TABLE
% \usepackage{graphicx}
\begin{table}[hb!]
\centering
\caption{A list of Sysmon events and their descriptions observed throughout the experiment}
\label{tab:table-sysmon}
\resizebox{0.55\linewidth}{!}{%
\begin{tabular}{|c|l|l|}
\hline
\textbf{Event} &
  \textbf{Event name} &
  \textbf{Description} \\ \hline
1 &
  Process creation &
  Full details about a newly created process \\ \hline
2 &
  \begin{tabular}[c]{@{}l@{}}A process changed a file \\ creation time\end{tabular} &
  \begin{tabular}[c]{@{}l@{}}Triggered when a file creation time is \\ modified by a process\end{tabular} \\ \hline
3 &
  Network connection &
  TCP/UDP connections on the machine \\ \hline
5 &
  Process terminated &
  \begin{tabular}[c]{@{}l@{}}Triggered when a process is terminated \\ on the machine\end{tabular} \\ \hline
7 &
  Image loaded &
  \begin{tabular}[c]{@{}l@{}}Triggered when a module (dd) is loaded\\ in a specific process\end{tabular} \\ \hline
8 &
  CreateRemoteThread &
  \begin{tabular}[c]{@{}l@{}}Triggered when hiding techniques are \\ detected such as process hollowing\end{tabular} \\ \hline
10 &
  ProcessAccess &
  \begin{tabular}[c]{@{}l@{}}Reported when a process opens another \\ process\end{tabular} \\ \hline
11 &
  FileCreate &
  \begin{tabular}[c]{@{}l@{}}Logged when a file is created on the \\ system or overwritten\end{tabular} \\ \hline
12 &
  \begin{tabular}[c]{@{}l@{}}RegistryEvent (Object \\ create and delete)\end{tabular} &
  \begin{tabular}[c]{@{}l@{}}Triggered when a registry key is created \\ or deleted\end{tabular} \\ \hline
13 &
  RegistryEvent (Value set) &
  \begin{tabular}[c]{@{}l@{}}Logged when values in the systems \\ registry are modified\end{tabular} \\ \hline
17 &
  PipeEvent (Pipe created) &
  Generated when a named pipe is created \\ \hline
22 &
  DNSEvent (DNS query) &
  \begin{tabular}[c]{@{}l@{}}Reported when a process executes a \\ DNS query\end{tabular} \\ \hline
23 &
  \begin{tabular}[c]{@{}l@{}}FileDelete (file delete \\ archived)\end{tabular} &
  Logged when a file is deleted \\ \hline
25 &
  \begin{tabular}[c]{@{}l@{}}ProcessTampering \\ (process image change)\end{tabular} &
  \begin{tabular}[c]{@{}l@{}}Triggered when hiding techniques are \\ detected such as process hollowing\end{tabular} \\ \hline
\end{tabular}%
}
\end{table}
% --------------------------------------

\subsection{Dynamic detection using a stream of Sysmon logs}
Employing dynamic detection mitigates the vulnerabilities of static detection, such as obfuscation and polymorphism \cite{or2019dynamic}. Subsequently, researchers have explored a broad spectrum of dynamic features to ascertain ransomware behaviour. Native Windows events, such as API calls and Opcodes, have been used throughout the literature to dynamically divulge ransomware behaviour, albeit with challenges such as API spoofing and limited system visibility \cite{ispahany2024ransomware}. Conversely, Sysmon operates at the executive level of the O/S (making techniques such as API spoofing more cumbersome) and is scalable, reliable, easy to install, and resilient to reboots. Importantly, Sysmon provides an avenue to forward system logs to a central logging platform without using well-known analysis tools that are easily detected by contemporary malware.
For this reason, we leverage Sysmon to supply a steady stream of security events to classify ransomware behaviour. Although custom rules can be configured, we propose using the default events to measure the efficacy of Sysmon events as a feature. By default, Sysmon offers 29 inbuilt events to indicate suspicious behaviour, some of which can be seen in Table \ref{tab:table-sysmon}. Importantly, although Sysmon discloses security-related events, not all originate from malicious software, and legitimate applications may trigger Sysmon events. 

\subsection{Dataset}

\subsubsection{The adopted approach for harvesting Sysmon events}

\begin{figure}
    \centering
    \includegraphics[width=0.7\linewidth]{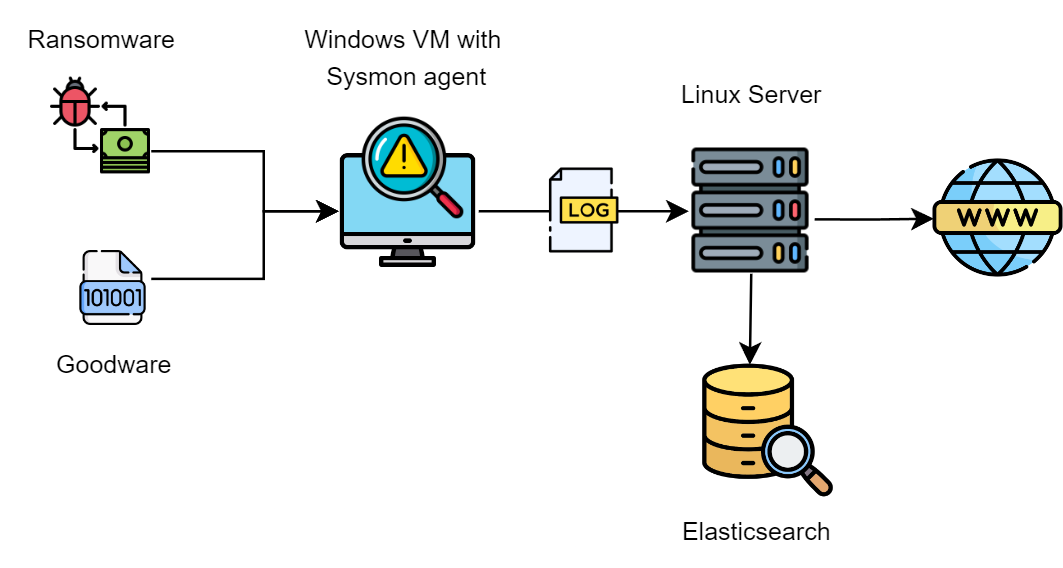}
    \caption{The laboratory setup used to collect both ransomware and goodware events to build the dataset.}
    \label{fig:lab-setup}
\end{figure}

To assess the system's effectiveness in detecting emerging ransomware without requiring a complete model rebuild, we created an imbalanced dataset of benign software (goodware) and ransomware events. This dataset focused on six prominent ransomware families and was generated using the lab setup shown in Fig. \ref{fig:lab-setup} which is discussed further in section. \ref{sec:lab-setup}. The data acquisition involved recording Sysmon events from the initiation of a program until the display of a ransomware demand note. However, generating a seamless sequence of non-malicious and malicious events presented a significant challenge. Each ransomware sample executed left the testing virtual machine in disrepair, necessitating a system restore to an earlier snapshot. Consequently, we combined Sysmon logs from both goodware and ransomware to form a cohesive training and testing environment for our models.\\
We sourced modern ransomware variants from online security platforms like VirusTotal and HybridAnalysis to stay abreast of cybersecurity threats. These variants included AvosLocker, BlackBasta, Conti, Hive, Lockbit, and REvil, chosen for their notoriety and prevalence in recent cyber attacks. Over 20,000 Sysmon event logs were collected from these activities across 50 ransomware attacks.\\
Each ransomware instance was run for a uniform duration of five minutes. Typically, the ransom note appeared before the five-minute mark. Following the launch of a ransomware sample, Sysmon began transmitting logs to a central repository until the ransom note materialised and the virtual machine ceased to function normally, preventing further event logging.\\
In real-world scenarios, benign events tend to outnumber malicious ones. Considering this, we deliberately created a dataset with more benign events. We obtained these events from various sources, including games and software from PortableApps, resulting in more than 176,000 benign events.\\
Our final dataset contained nearly 200,000 events, a mix of ransomware and benign activities interspersed throughout sequences of ransomware events. By periodically introducing new ransomware families into the dataset, we created conditions that mimic the evolutionary nature of ransomware threats. This approach allowed us to assess our models' adaptability to the continuous changes seen in ransomware attacks. 

%\----------------------------------------------------------
% Figure: Lab setup
%\----------------------------------------------------------

%\----------------------------------------------------------

\subsubsection{Dataset preparation}\label{dataset-preparation}

In our methodology for preparing the dataset for analysis, we employed the Standard Scaler algorithm to normalise the data, ensuring a consistent scale across features. The Standard Scaler function is pivotal in transforming each feature to have a mean value of zero and a standard deviation of one. This normalisation is accomplished by subtracting the mean value of each feature from individual data points and then dividing the result by the standard deviation of that feature \cite{thara2019auto}. By doing so, the transformed dataset enhances the performance of many machine learning algorithms by treating all features equally. This is especially important when the feature set contains different units (for example, the difference between centimetres and kilograms). This approach not only simplifies the computational process but also prevents features with larger scales from dominating those with smaller scales, thus maintaining a balanced influence of all features in the predictive modelling process.
The Standard Scaler is calculated by the following:

\begin{align}
     \hat{x} = \frac{x - \mu}{\sigma}
\end{align}

Where $\hat{x}$ equals the normalised feature value, $x$ is the feature value, $\mu$ is the mean and $\sigma$ is the standard deviation. 

In the initial phase of our study, we prepared our model using the first 40,000 events from our dataset as the training set. In contrast, the subsequent events were segmented into additional batches of 10,000 for ongoing training. The distribution of benign and ransomware events across each batch is detailed in Table. \ref{tab:spread}. Given the significant class imbalance within our dataset, we incorporated the Synthetic Minority Oversampling Technique (SMOTE) to enhance the representation of the minority class in the training dataset. SMOTE is a widely acknowledged method to mitigate issues arising from imbalanced data sets by synthetically generating new examples in the minority class \cite{lusa2012evaluation}. This technique is particularly noted for its effectiveness in contexts involving low-dimensional data. However, its efficacy may diminish in high-dimensional spaces due to the complexity of accurately generating synthetic examples in such environments. Despite these limitations, we employ SMOTE due to the limited number of features selected, as explained in section \ref{sec:selection}. Overall, SMOTE generally has an improved classification quality by providing a more balanced dataset, facilitating more effective learning and generalisation for the underlying model.

% --------------------------------------
% DATASET TABLE
\begin{table}[]
\centering
\caption{The distribution of benign and ransomware events within the dataset. The initial model is built using the training batch and then updated with subsequent batches containing ransomware and benign events}
\label{tab:spread}
\begin{tabular}{|c|c|c|}
\hline
\textbf{Batch}                                           & \textbf{\begin{tabular}[c]{@{}c@{}}Benign\\ Events\end{tabular}} & \textbf{\begin{tabular}[c]{@{}c@{}}Ransomware\\ events\end{tabular}} \\ \hline
\begin{tabular}[c]{@{}c@{}}Training\\ batch\end{tabular} & 38,817                                                           & 1,181                                                                \\ \hline
1                                                        & 9,747                                                            & 254                                                                  \\ \hline
2                                                        & 9,102                                                            & 899                                                                  \\ \hline
3                                                        & 9,588                                                            & 413                                                                  \\ \hline
4                                                        & 9,397                                                            & 607                                                                  \\ \hline
5                                                        & 8,407                                                            & 1,594                                                                \\ \hline
6                                                        & 9,760                                                            & 241                                                                  \\ \hline
7                                                        & 9,826                                                            & 175                                                                  \\ \hline
8                                                        & 7,463                                                            & 2,538                                                                \\ \hline
9                                                        & 8,818                                                            & 1,183                                                                \\ \hline
10                                                       & 6,876                                                            & 3,125                                                                \\ \hline
11                                                       & 8,340                                                            & 1,661                                                                \\ \hline
12                                                       & 6,057                                                            & 3,844                                                                \\ \hline
13                                                       & 9,895                                                            & 106                                                                  \\ \hline
14                                                       & 9,443                                                            & 558                                                                  \\ \hline
15                                                       & 9,301                                                            & 700                                                                  \\ \hline
16                                                       & 5,305                                                            & 1,536                                                                \\ \hline
\textbf{Total}                                           & \textbf{176,130}                                                 & \textbf{20,710}                                                      \\ \hline
\end{tabular}
\end{table}
% --------------------------------------

\subsection{Feature extraction and selection}

\subsubsection{Feature extraction using fastText}

\begin{figure}
    \centering
    \includegraphics[width=.35\linewidth]{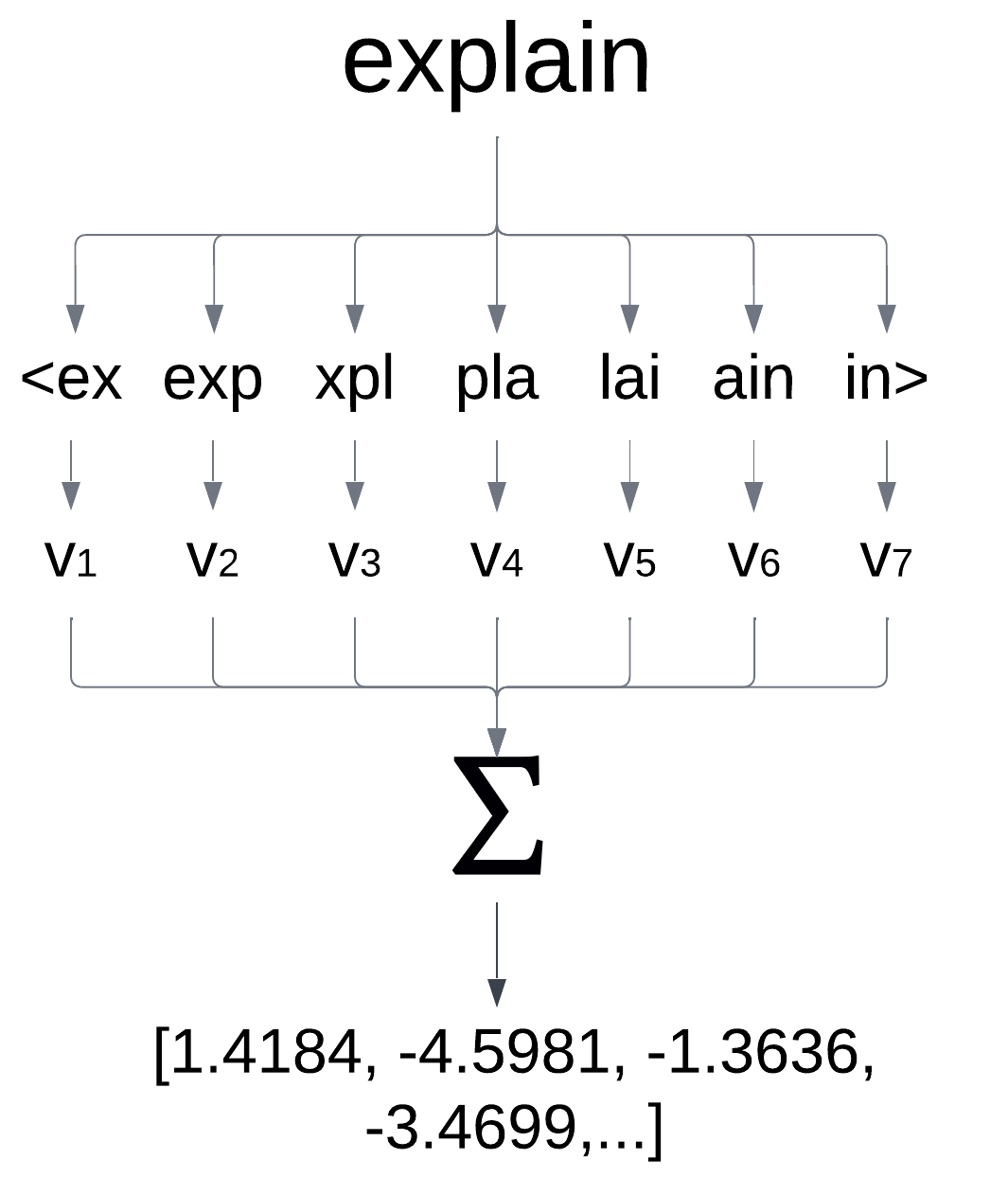}
    \caption{fastText conversion to vectors using n-grams. The above example shows the process of converting the word "explain" into vectors where $v_n$ represents the vector representation of the corresponding $n$-gram}
    \label{fig:fasttext}
\end{figure}

Our system employs fastText to transform words into vector representations for feature extraction. Unlike other embeddings, such as Word2vec, that treat words as indivisible entities and neglect their sub-structural components, fastText considers the internal structure of words. Specifically, as illustrated in Fig. \ref{fig:fasttext}, fastText decomposes words into character $n$-grams and computes a word's vector by summing the vectors of its constituent $n$-grams. This approach enhances the model's performance, particularly with complex and rare words, by providing a more nuanced representation \cite{bojanowski2017enriching}.

To elaborate, let's denote the set of $n$-grams for a given word by $\mathcal{G}_{w}$, a subset of the dictionary of all $n$-grams of size $G$. Each $n$-gram, $g$, from this set is mapped to a unique vector $\textbf{z}_g$ in an $N$-dimensional space. Similarly, vectors $\textbf{v}_c$ in an $N$-dimensional space represent the context words surrounding the target word. With these vectors in place, we define a scoring function that evaluates the alignment between the target word and its context within the embedding space.

\begin{align}
  s(w, c) =  \sum_{g \in \mathcal{G}_{w}}^{|G|} \bigg ( \textbf{z}^T_g\textbf{v}_c \bigg ) = 1 
  \label{eqn:SF}
\end{align}

\subsubsection{Feature selection technique}\label{sec:selection}

Fifty-two features are extracted from each Sysmon event; however, not all of these features significantly contribute to achieving high accuracy in ransomware detection. In our study, the Pearson Correlation Coefficient (PCC) is a critical tool for identifying the most significant features within our dataset, particularly those that strongly correlate with the target class. The PCC is a statistical measure of the linear correlation between each feature $x$ and the target class $y$, providing a value that indicates the strength and direction of the association. This value is derived from the normalised form of the covariance between the two variables, where the normalisation ensures that the coefficient remains between -1 and +1, with values closer to these extremes indicating stronger relationships. Assuming $\overline{x}$ and $\overline{y}$ are the averages of $x$ and $y$, the PCC can be calculated using the following formula \cite{venkatesh2019review}:

\begin{align}
     PCC(x,y) = \frac{\sum(x - \overline{x})(y - \overline{y})}{\sqrt{\sum(x - \overline{x})^2(y - \overline{y})^2}}
\end{align}

\begin{figure}
    \centering
    \includegraphics[width=0.75\linewidth]{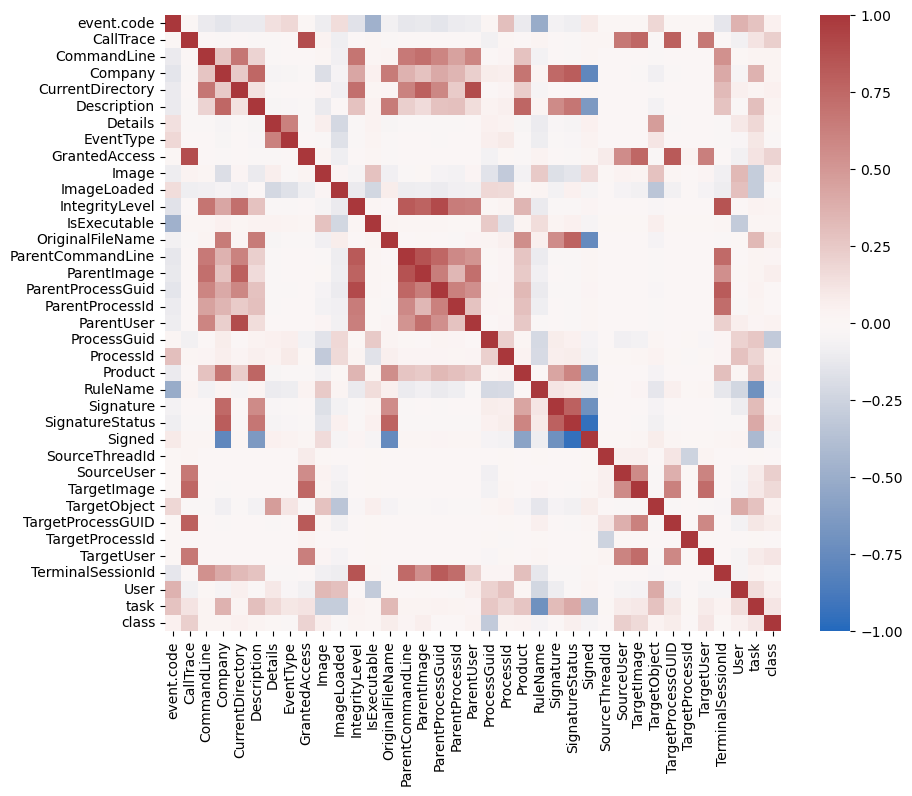}
    \caption{A heatmap showing the correlations between features within the training dataset}
    \label{fig:heatmap}
\end{figure}

We computed the Pearson Correlation Coefficient (PCC) across the training dataset and displayed the results as a heatmap in Fig \ref{fig:heatmap}. Our analysis revealed positive correlations among several features. Consequently, we identified the most significant features that exhibit strong positive correlations with the target class value, such as CallTrace, GrantedAccess, SourceUser, TargetImage, TargetUser, and Task. Detailed descriptions of these features are provided in Table. \ref{tab:important-features}.

% --------------------------------------
% TABLE: Sysmon features table
% --------------------------------------
\begin{table}[]
\centering
\caption{The most significant features throughout the training dataset as determined by the PCC}
\label{tab:important-features}
\resizebox{0.6\linewidth}{!}{%
\begin{tabular}{|l|l|}
\hline
\textbf{Feature} & \textbf{Description}                                                                                                                                          \\ \hline
CallTrace        & \begin{tabular}[c]{@{}l@{}}Stack trace of where the open process is called. Included is the \\ DLL and the relative virtual address of the functions\end{tabular} \\ \hline
GrantedAccess    & \begin{tabular}[c]{@{}l@{}}The access flags (bitmask) associated with the process rights \\ requested for the target process\end{tabular}                     \\ \hline
SourceUser       & Name of the account that runs the source process.                                                                                                             \\ \hline
TargetImage      & File path of the executable of the target process                                                                                                             \\ \hline
TargetUser       & \begin{tabular}[c]{@{}l@{}}Name of the account that runs the targeted process which \\is accessed  \end{tabular}                                                                                        \\ \hline
Task             & \begin{tabular}[c]{@{}l@{}}A description of the Sysmon event called (example process \\created)\end{tabular}                                                                                                                                                            \\ \hline
\end{tabular}%
}
\end{table}
% --------------------------------------

The feature selection engine in our ransomware detection system periodically processes incoming data and calculates the significance of each feature using the PCC. This step is crucial as the value of the target class, which is essential for calculating the PCC, is unavailable during the prediction phase. By periodically recalculating the PCC with new batches of data, we ensure that our feature selection remains up-to-date and relevant to the evolving dynamics of the dataset. This systematic approach allows us to streamline our model by focusing on features that show significant correlations with the target class, thereby enhancing our analytical models' predictive accuracy and efficiency. If changes occur in the feature set, the model is retrained using historical data. Addressing this limitation is planned for future work.

\newpage
\subsection{The proposed classification model for ransomware detection}

\subsubsection{The CNN model}

Our experiment employs multiple one-dimensional convolutional neural networks (1D CNNs) to process and analyse data. Specifically, the input for our models consists of a stream of Sysmon event data, which is inherently sequential and well-suited for a 1D CNN. This architecture is particularly adept at handling such data formats, as it can efficiently capture the temporal dependencies and patterns within the continuous data stream. By utilising multiple 1D CNNs, we aim to enhance the model's ability to extract and learn various discriminative features from the Sysmon event data, thereby improving the accuracy and robustness of the detection system.
Convolutional Neural Networks (CNNs), particularly one-dimensional (1D) CNNs, are an advanced class of neural networks that diverge from traditional fully connected networks by employing convolution operations instead of matrix multiplications in their hidden layers \cite{goodfellow2016deep}. In 1D CNNs, these convolutions are performed along the input data using kernels or filters of predefined widths and strides. Each filter systematically slides across the input data, capturing local dependencies and spatial hierarchies by applying the same filter across different input parts, thus preserving the spatial relationship between points \cite{stankovic2023convolutional}. This architecture allows 1D CNNs to excel in applications where data can be represented in temporal or spatial dimensions, such as audio signals, time-series data, or any sequential data. The convolutional layers are particularly adept at focusing on local patterns within the data, enabling the network to detect complex features at various scales and depths. This capability makes 1D CNNs highly effective for binary classification tasks, particularly ransomware detection tasks, where distinguishing between classes can hinge on recognising subtle, localised patterns in a dataset. By leveraging localised feature detection, 1D CNNs provide a powerful binary classification tool, offering high accuracy and efficiency in processing sequential data.
The operation of a one-dimensional convolutional neural network (1D CNN) can be expressed mathematically using the following:

Assuming $x[n]$ denotes the input data, $w[m]$ represents the weights of the filter's kernel for each filter element indexed by $m$, $b$ is the bias term associated with the filter, and $n$ is the index used for computing the output signal at each step of the convolution process. If the convolution is applied with one stride, then the output $y$ is given by Formula. \ref{form:cnn}:

\begin{equation}\label{form:cnn}
     y[n] = b + \sum_{m=0}^{M-1} w[m] \cdot x[n + m]
\end{equation}

$M$ is the total number of elements in the convolution filter, and $x[n + m]$ refers to the input data at the position shifted by $m$ from the current position $n$ in the output signal. After the convolution process, an activation function is usually applied to introduce non-linearity using Formula. \ref{form:activation}:

\begin{equation}\label{form:activation}
     a_k(n) = \sigma(y_k(n))
\end{equation}

Where $\sigma$ is the activation operation. In summary, the output of the 1D CNN is given by the convolution operation followed by an activation function as denoted in Formula. \ref{form:full-cnn}:

\begin{equation}\label{form:full-cnn}
     a_k[n] = \sigma(b + \sum_{m=0}^{M-1} w[m] \cdot x[n + m])
\end{equation}

This process occurs for every filter in the convolution layer, allowing the CNN to learn different features from the input signal.

\subsubsection{The LSTM model}

\begin{figure}
    \centering
    \includegraphics[width=0.60\linewidth]{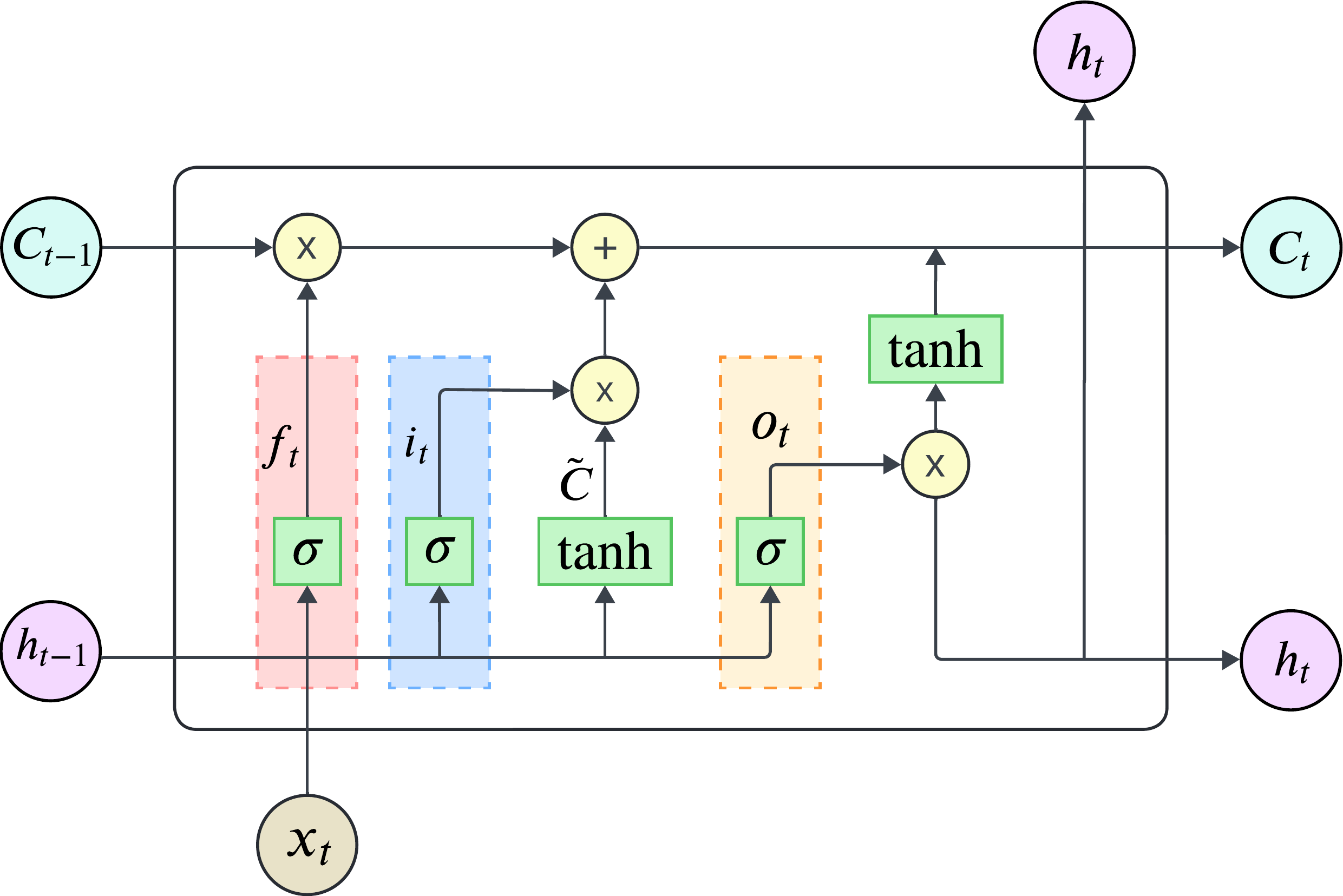}
    \caption{The structure of an LSTM model. The red region denoted by $f_t$ outlines the forget gate; The blue shaded region denoted by $i_t$ represents the input gate, and the orange region ($O_t$) represents the output gate}
    \label{fig:lstm-model}
\end{figure}

The Long Short-Term Memory (LSTM) model, a specialised type of recurrent neural network (RNN) was designed to overcome the vanishing gradient problem that plagues standard RNNs \cite{hochreiter1997long}. Unlike traditional RNNs, which lack mechanisms to retain information over extended sequences, leading to issues such as vanishing gradients, LSTMs incorporate specialised memory cells. These cells are equipped with input, output, and forget gates that meticulously regulate information flow, enabling the preservation of data across lengthy intervals. Additionally, LSTMs utilise constant error carousels (CECs) to maintain a stable gradient during training, which prevents gradients from vanishing or exploding and supports effective learning over time lags of more than 1000 steps, even amidst noisy inputs. This structural complexity allows LSTMs to excel in processing long sequences and complex patterns, outperforming standard RNNs in numerous applications. This intricate gating mechanism enables LSTMs to excel at tasks that require the handling of sequential data, such as time-series analysis, by maintaining essential historical information over long durations. Such capabilities make LSTMs particularly valuable in fields where patterns across time are critical for prediction and analysis.

Fig. \ref{fig:lstm-model} presents the structure of an LSTM model. The operations within the LSTM unit for each time step can be calculated by first determining what information is discarded using the Formula. \ref{eqn:forget-gate}:
\begin{equation}\label{eqn:forget-gate}
    f_t = \sigma(W_f \cdot [h_{t-1},x_t] + b_f)
\end{equation}
Where $f_t$ is the forget gates activation vector, $\sigma$ denotes the sigmoid function, $W_f$ represents the weights of the forget gate, $h_{t-1}$ is the previous output, $x_t$ is the input at time step $t$ and $b_f$ is the forget gates bias.
The input gate then updates the cell state by regulating the flow of new information into the cell, which is determined by Formula. \ref{eqn:input-gate} and \ref{eqn:cell-update-1}:
\begin{equation}\label{eqn:input-gate}
    i_t = \sigma(W_i \cdot [h_{t-1},x_t] + b_i)
\end{equation}

\begin{equation}\label{eqn:cell-update-1}
    \tilde{C_t} = \tanh(W_C \cdot [h_{t-1},x_t] + b_C)
\end{equation}

Where $i_t$ is the input gates activation vector, $\tilde{C_t}$ is the candidate values, calculated by a tanh layer, to add to the state, $W_i$ and $W_C$ are the weights, and $b_i$ and $b_C$ are the biases. Next, the old cell state $C_{t-1}$ is updated with the new cell state $C_t$, using the output generated from the forget and input gates. This operation discards or retains information from the previous cell state and mixes the new candidate values gated by the input gate, denoted by Formula. \ref{eqn:cell-update-2}:

\begin{equation}\label{eqn:cell-update-2}
    C_t = f_t*C_{t-1}+i_t*\tilde{C}
\end{equation}

The output gate (Formula \ref{eqn:output-gate-1} is determined after receiving $h_{t-1}$ and $x_t$ as input at time $t$, where $W_o$ represents the weights and $b_o$ is the bias:

\begin{equation}\label{eqn:output-gate-1}
    o_t = \sigma(W_o \cdot[h_{t-1},x_t] + b_o)
\end{equation}

Finally, the output of the LSTM $h_t$ is determined from the output gate vector $o_t$ and the activation of the cell state $C_t$ using Formula \ref{eqn:output-gate-2}: 

\begin{equation}\label{eqn:output-gate-2}
    h_t = o_t*\tanh(C_t)
\end{equation}

\subsubsection{Attention mechanism}

%\----------------------------------------------------------
% Figure: Attention mechanism
%\----------------------------------------------------------
\begin{figure}
    \centering
    \includegraphics[width=0.60\linewidth]{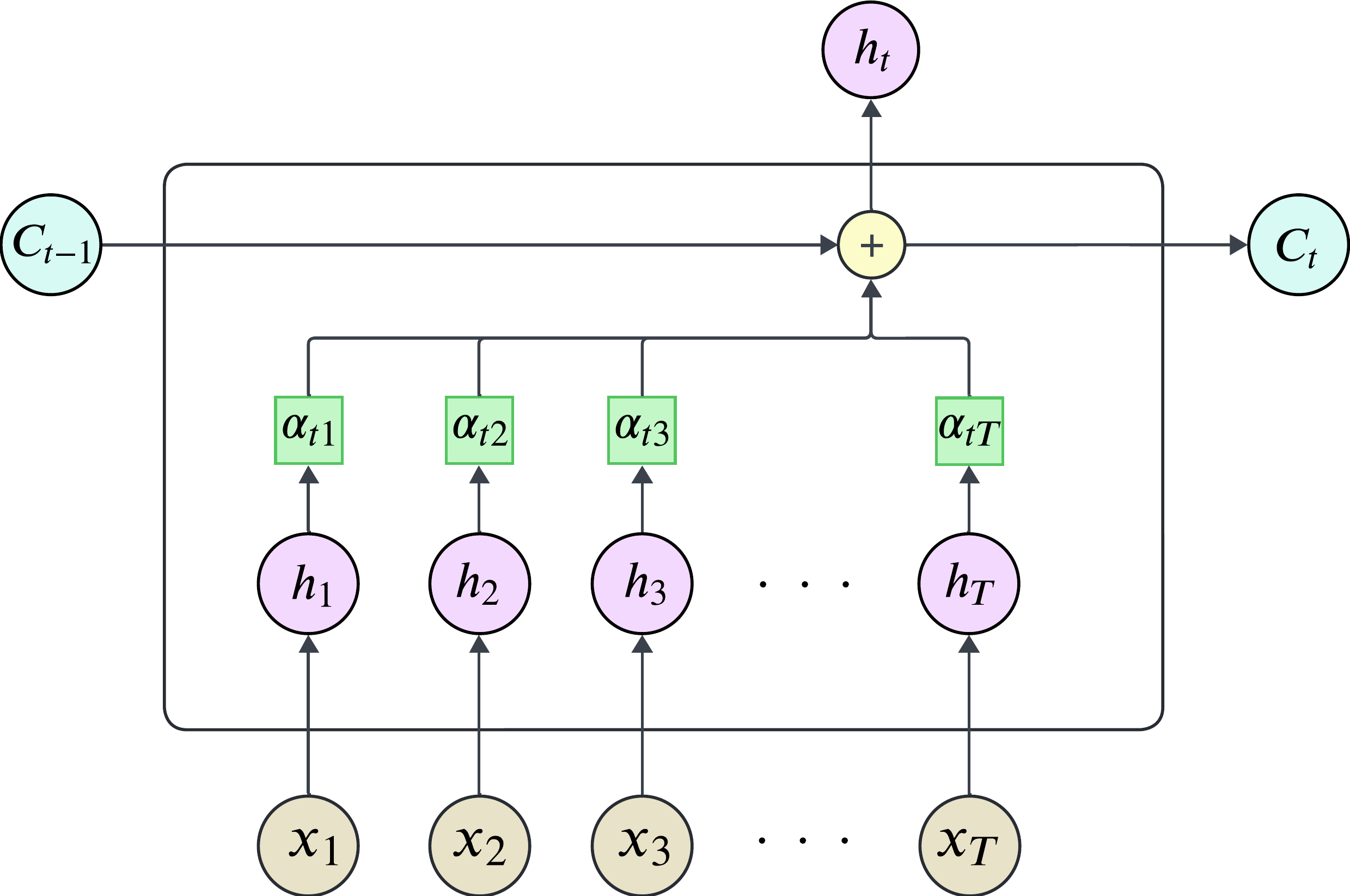}
    \caption{The attention mechanism}
    \label{fig:attention}
\end{figure}
%\----------------------------------------------------------

Attention mechanisms have recently garnered significant interest for their effectiveness in time series forecasting tasks across various domains outside ransomware detection \cite{zheng2021understanding, chen2019exploring, zhang2018attention}. These mechanisms enhance forecasting performance by capturing the dynamic influence of exogenous series most relevant to the target series. In LSTM networks, attention mechanisms address the limitations of traditional LSTMs, which tend to treat all parts of the input sequence equally when updating the hidden state. This equal treatment can hinder the model's ability to differentiate between more and less relevant parts of the sequence, potentially limiting its performance—especially when only specific input segments contain critical information for accurate predictions. Attention mechanisms overcome this limitation by focusing on the most relevant parts of the input sequence, assigning higher attention weights to these crucial elements. This prioritisation enables the model to emphasise the most critical information for the task, leading to improved classification performance. 

The following method can calculate the attention mechanism shown in Fig. \ref{fig:attention} to determine which parts of the LSTM's output are the most relevant. First, using Formula \ref{eqn:attention}, each hidden state is given a score to measure its importance relative to the current output:

\begin{equation}\label{eqn:attention}
     e_t = v_a^\top \tanh(W_a \cdot h_t)
\end{equation}

Where $h_t$ is the hidden state output from the LSTM at time step $t$, and $C_t$ is the memory cell state from the encoder unit. $W_a$ represents the weight matrix that transforms the hidden state $h_t$, and $v_a$ is a weight vector that determines the transformed hidden state's importance. Next, the scores are converted into attention weights using the following softmax function to determine the contribution of each hidden state to the final context vector using Formula. \ref{eqn:contextvector}:

\begin{equation}\label{eqn:attention2}
     \alpha_t = \frac{\exp(e_t)}{\sum_{k=1}^{T}\exp(e_k)}
\end{equation}

Where $\alpha_t$ represents the attention weight for the hidden state $h_t$ and $T$ represents all time steps across the hidden states of the LSTM module. Once the attention weights are determined, the context vector $c$ can be calculated using the following formula:

\begin{equation}\label{eqn:contextvector}
     c = \sum_{t=1}^{T}\alpha_t h_t
\end{equation}

Finally, the final output can be determined by combining the context vector $c$ with the hidden state of the current time step using the following formula:

\begin{equation}\label{eqn:attention2}
     output = f(c,h_T)
\end{equation}

\subsubsection{The combined CNN-LSTM models}

%\----------------------------------------------------------
% FIGURE: Proposed models
%\----------------------------------------------------------

\begin{figure}
    \centering
    \includegraphics[width=0.95\linewidth]{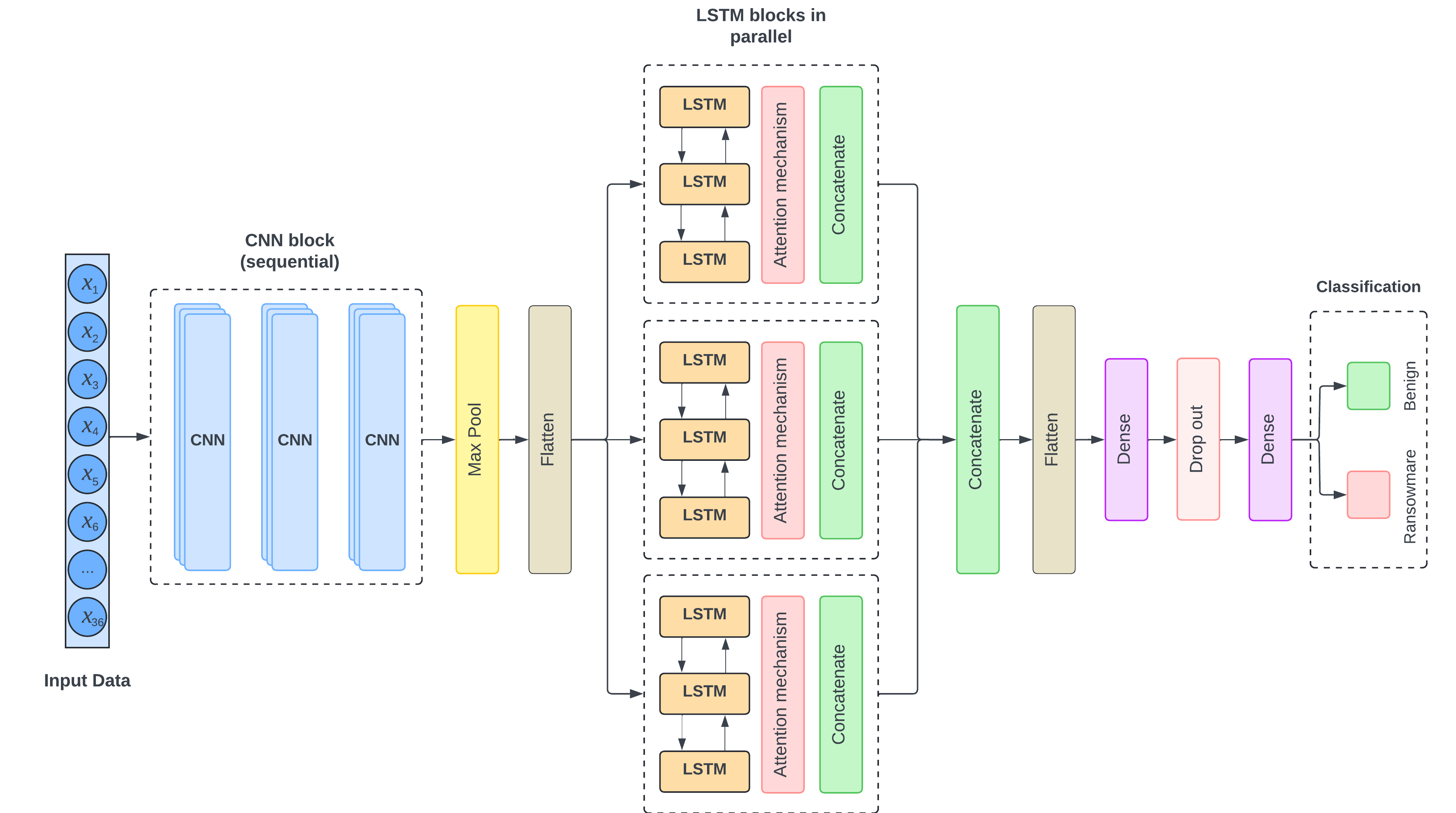}
    \caption{The proposed CNN-LSTM model, comprising of parallel LSTM blocks and attention mechanisms}
    \label{fig:model-arch}
\end{figure}
%\----------------------------------------------------------

In Fig. \ref{fig:model-arch}, we present our CNN-LSTM ransomware detection classifier, which effectively combines convolutional neural networks (CNNs) and long short-term memory networks (LSTMs) to detect ransomware activities within a stream of Sysmon event data. We propose a model configuration that integrates stacked CNN and LSTM models operating in parallel. This setup allows the model to capture different data characteristics: the CNN modules extract spatial features from the input data, while the LSTM modules sequence through the data to capture temporal dependencies. We run the LSTM modules parallel to alleviate the time constraints typically associated with processing LSTMs \cite{dai2019grow}. A widely adopted variant of this architecture in the literature involves sequentially "stacking" LSTM modules after CNN modules \cite{zhang2020deepmal, luan2019research, ozkok2022hybrid, lu2020cnn, akhtar2022detection}. This configuration allows for streamlined temporal analysis by processing sequentially refined features. However, stacking LSTMs in this manner may introduce delays in the training and testing phases. The LSTM layers can only commence processing after the complete execution of the CNN layers, and their sequential operation forms a computational bottleneck that potentially prolongs the temporal analysis. This arrangement organises data processing in a logical sequence but may encounter challenges related to efficiency and speed. This is particularly problematic in ransomware detection within real-time data streams, where time efficiency is critical. For this reason, we propose the LSTM modules run in parallel to help ease this bottleneck.

The outputs from the LSTM modules are then passed through an attention mechanism, which helps the model focus on the most informative features before passing through the classification layers. This approach enhances the performance of the LSTM models and is particularly valuable for ransomware detection, where specific signal patterns can be more indicative of malicious activity. The output of these layers is then merged using a concatenation layer before the output is flattened. The flattened output from the LSTM blocks ensures that the input to the dense layers, which make the final classification decision, is compact yet rich with the most critical information distilled from earlier layers. We employ Dropout as a regularisation technique to mitigate overfitting in our system \cite{srivastava2014dropout}, effectively preventing complex co-adaptations on the training data \cite{hinton2012improving}. Dropout randomly omits a subset of neurons during the forward training pass, making them inactive. Consequently, these neurons do not participate in the forward pass or receive weight updates during the backward pass. This randomness helps to ensure that no single neuron becomes overly dependent on the specific features of other neurons, thereby enhancing the model's generalisation capability.
Finally, the architecture concludes with dense layers that output the final classification into ransomware or benign categories, demonstrating high efficiency and effectiveness in preliminary tests. This approach makes the classifier robust against ransomware signatures' complex and dynamic nature.

%\----------------------------------------------------------
% FIGURE: Mini-batch updates
%\----------------------------------------------------------
\subsection{Updating the model using mini-batches}

\begin{figure}
    \centering
    \includegraphics[width=.80\linewidth]{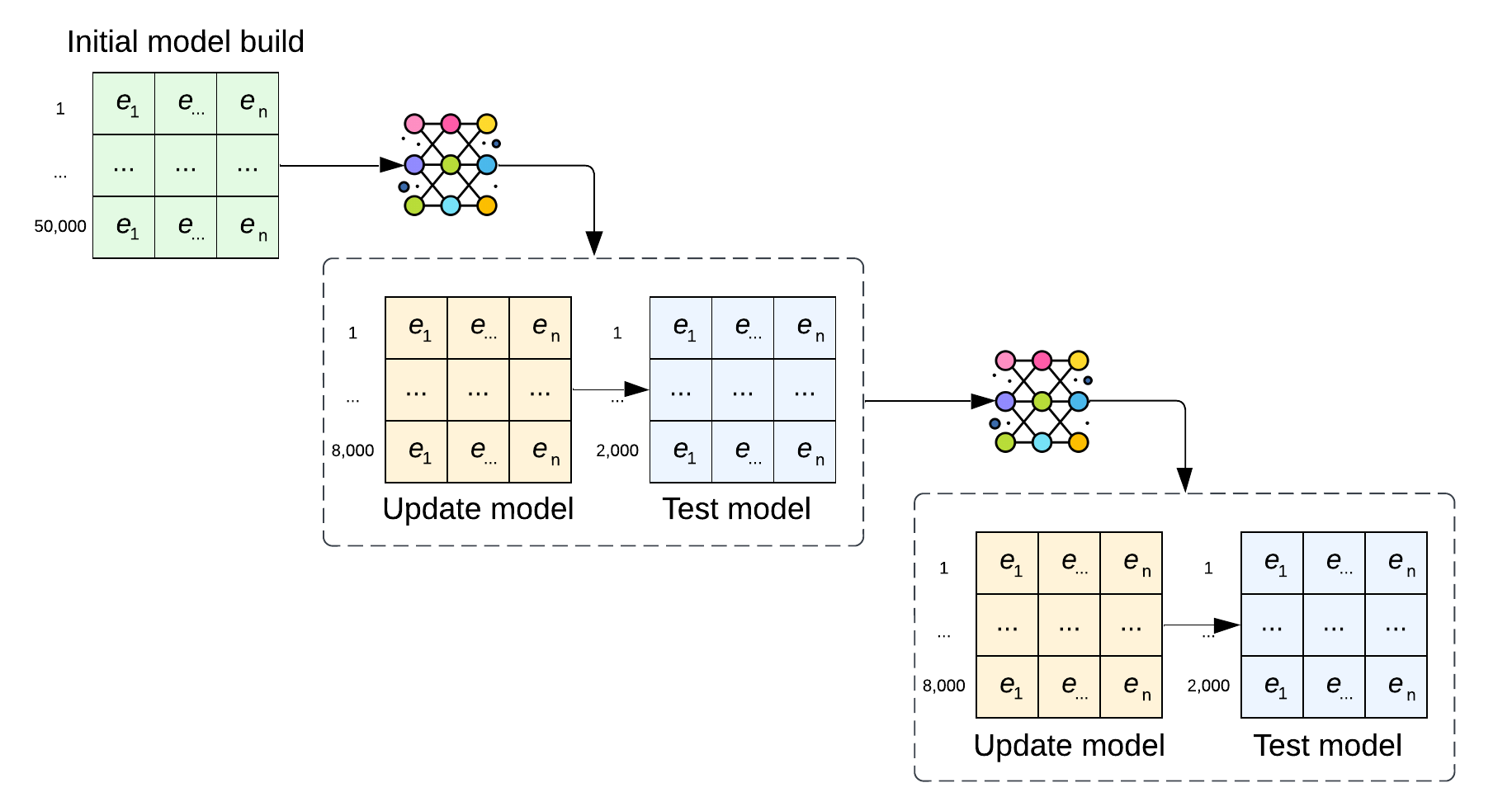}
    \caption{The process to incrementally update the CNN-LSTM model. The initial model is built using training data. Subsequent mini-batches of 10,000 events are used to update the model weights periodically}
    \label{fig:model-updates}
\end{figure}
%\----------------------------------------------------------

Fig. \ref{fig:model-updates} outlines the methodology for constructing and refining our model. Initially, the model undergoes training and validation with 40,000 events. Following its deployment, we adopt an incremental learning approach, where the model is updated via batches, each consisting of 10,000 events. We have divided the data into 15 batches to update the model progressively after the initial training phase. To evaluate the model's capacity for identifying previously unknown ransomware, each update incorporates novel samples from various ransomware families not included in the original training dataset. To assess the model's ability to classify events accurately, each batch used for updating the model is divided into an 80:20 ratio. This means that 80\% of the events (or 8,000 events) in each batch are allocated for training, while the remaining 20\% are used to test the model's performance. The process of updating the model can be seen in the Algorithm. \ref{alg:cap}

%\----------------------------------------------------------
% ALGORITHM 1
%\----------------------------------------------------------
\begin{algorithm*}

\caption{iCNN-LSTM \textbf{Symbols:} $E$: The Sysmon event, $x$: training feature values, $y$: target class values, $W$: Window of values (mini-batch), $M$: The iCNN-LSTM model}\label{alg:cap}
\begin{algorithmic}
\Function{}{}
    \State Initialise empty batch window $W$
    \State Initialise empty DataFrame 
    \State Set batch\_size to 1024

    \While{True}
        \State $E$ $\gets$ Next Sysmon event \Comment{Select the next Sysmon event in the data stream}
        \If{$E$ is not None:}
            \State $W \gets E$ \Comment{Ingest the event into the mini-batch }
        \EndIf

        \If{$W$ = length of batch\_size}
            \If{No model ($M$) exists}
                \State Create model $M$  
            \Else
                \State Load model $M$ \Comment{Load previous model if exists}
                \State Freeze weights in $M$ \Comment{Freeze the model weights to update the outer layers}
            \EndIf
            \For{$E \in W$}
                \For{All \textbf{$x$}, $y \in E$}
                    \State $\bar{x} \gets $ Convert $x$ to word embedding \Comment{Convert  to vector using fastText}
                    \State Append $\bar{x}$ and $y$ to DataFrame 
                \EndFor
            \EndFor
            \State $\hat{x}, \hat{y} \gets$ Pre\_process\_data(DataFrame)
            \State (train\_$x$, train\_$x$, test\_$x$, test\_$y$) $\gets$ Split data for training and testing $M$ ($\hat{x}, \hat{y}$) 
            \State Fit and train $M$ (train\_$x$, train\_$y$) \Comment{Update the model using the training data}
            \State Evaluate model $M$ (test\_$x$, test\_$y$) \Comment{Evaluate the model using the testing data}
            \State Save the model in $M$
        \EndIf
    \EndWhile
\EndFunction

\Function{Pre\_process\_data}{}
    \For{All values $\in$ DataFrame}
        \State $\Ddot{x} \gets$ All feature values \Comment{Select all feature values in the Dataset up until the class value}
        \State $\Ddot{y} \gets$ All class values \Comment{Select the class value}
        \State $\Ddot{x} \gets$ Calculate PCC for all $\Ddot{x}$ \Comment{Calculate the Pearson Correlation Coefficient (PCC)}
        \State Select top features in DataFrame \Comment{Select the highest performing features based on their PCC value}
        \State return $(\Ddot{x}, \Ddot{y})$
    \EndFor
\EndFunction

\end{algorithmic}
\end{algorithm*}

%\----------------------------------------------------------
\section{Experiment and results}\label{sec:experiment} 

\subsection{Lab setup}\label{sec:lab-setup}

As discussed in existing literature, a prevalent method for analysing ransomware behaviour involves detonating ransomware samples in sandbox environments like Cuckoo to collect logs after execution. However, this approach faces challenges as modern ransomware can often detect its environment and may withhold execution to avoid detection. To counter this, we have meticulously designed our laboratory environment to mimic a real-world production endpoint with full internet connectivity, as illustrated in Fig. \ref{fig:lab-setup}. This setup includes deploying isolated virtual machines equipped with Windows 11, 4GBs of RAM, four cores, and 80GBs of storage. While using bare metal for detonating ransomware is ideal, virtual machines offer the advantage of being swiftly restored to a previous state after infection.\\
To simulate a production environment better, we populated the file system with over 1000 dummy files in various formats, including CSV, GIF, JPEG, MP4, PNG, PPT, PDF, docx, and zip. These files were obtained from file samples \footnote{https://filesamples.com/}, enhancing the realism of our setup.\\
Additionally, Sysmon agents are installed on the Windows virtual machines within this network to collect and forward Sysmon events for classification, as shown in Fig. \ref{fig:architecture}. This setup facilitates effective ransomware behaviour analysis and enhances the realism of our testing environment to elicit more natural ransomware actions.
All Sysmon logs were forwarded to a cloud-based central logging server for further analysis, utilizing an Elasticsearch stack (version 8.9.0) with 45GB of storage, 1GB of RAM, and up to 8 vCPUs. The deep learning environment used to process the dataset and run the proposed iCNN-LSTM model was built with Python 3.9.18 and TensorFlow 2.17.0.

\subsection{Hyper-parameter tuning}

We used Optuna for automatic hyper-parameter training \cite{optuna_2019}. Optuna is a versatile tool for hyperparameter optimisation, widely recognised for its efficacy in refining machine learning models to enhance their performance. This open-source framework facilitates the optimisation process through advanced searching and pruning algorithms. These algorithms not only expedite the search for optimal hyperparameters but also enhance the cost-effectiveness of the process. Specifically, Optuna’s ability to efficiently prune suboptimal trials enables it to concentrate resources on more promising hyperparameter configurations. This strategic allocation significantly reduces computational waste, allowing for a faster and more efficient determination of the best parameters for any given model. This methodology streamlines model development and ensures the models are robust and tuned to their highest potential performance. The hyper-parameters producing the best overall performance can be seen in Table. \ref{tab:params}.

\begin{table}
    \centering
    \caption{Hyper-parameters chosen for the model}
    \label{tab:params}
    \begin{tabular}{|c|c|} \hline 
        \textbf{Parameter} & \textbf{Value} \\ \hline 
         CNN kernel size& 9\\ \hline 
         CNN filters& 32\\ \hline
         LSTM units&384\\\hline
         Dropout rate (LSTM)& 0.10326648213511579\\ \hline
         Activation type&tanh, Sigmoid\\\hline
         Optimizer&Adam optimiser\\\hline
         Learning rate&0.001\\ \hline
         Dense Layers& 80, 2\\ \hline
         Dropout rate (FC layer)& 0.4057318990206279\\ \hline
         Epochs&100\\\hline
         Batch size&1,024\\\hline
    \end{tabular}
\end{table}

\subsection{Evaluation metrics}

\bigskip

It is common practice to employ the Area Under the Receiver Operating Characteristic (AUC-ROC) curve to measure model performance, as widely documented throughout the literature. However, when applied to imbalanced datasets, the AUC-ROC can exhibit considerable variability and may provide misleadingly optimistic performance evaluations. This variability stems from the AUC-ROC's tendency to not adequately penalise false negatives in the minority class, which are often more critical in such contexts. In contrast, the F-measure, which represents the harmonic mean of the precision and recall, offers a significant advantage in scenarios with imbalanced datasets \cite{ispahany2024ransomware}. The F-measure is particularly valuable as it considers both false positives and false negatives, providing a more balanced and realistic evaluation of model performance where class distribution is skewed \cite{christen2023review}.
Since we focus on reducing ransomware attacks, misclassifying ransomware events as benign (false negatives) poses a more severe risk than incorrectly identifying benign software as ransomware (false positives). Consequently, our primary objective is to minimise false negatives to enhance the security and integrity of the systems we aim to protect. To achieve this, we focus on maximising the recall measure, which quantifies the proportion of actual positives that are correctly identified. This is implemented by selecting a lower threshold for the F-Measure, which inherently may increase false positives \cite{christen2023review}. While this adjustment can compromise precision by allowing more false positives, it strategically prioritises the reduction of false negatives, ensuring that fewer ransomware attacks go undetected. This methodological choice underscores our approach to balancing the need for high recall with acceptable levels of precision, reflecting a tailored response to the specific consequences of misclassifications in the context of ransomware detection.

Aligned with our emphasis on minimising false negatives in ransomware detection, we have opted for an F-measure configuration tailored to prioritise recall. To this end, we utilise an F-measure where the beta value is set to 2, denoted as F2-score. This setting shifts the emphasis towards recall by weighing it twice as heavily as precision. A beta value of 1 in the F-measure formula treats precision and recall as equally important. Conversely, a less than one beta value emphasises precision, making the measure more sensitive to false positives. The formula used to calculate the F2-score is as follows:

\bigskip
\begin{align}
    F_\beta = (1 + \beta^2) \cdot \frac{precision \cdot recall}{(\beta^2 \cdot precision) + recall}
\label{eqn:f2-score}
\end{align}
\bigskip

Where $\beta$ is the beta value, which determines the weight of precision and recall in the harmonic mean. The precision and recall can be calculated as follows:

\bigskip
\begin{align}
    precision = \frac{TP}{TP + FP}
\end{align}

\begin{align}
    recall = \frac{TP}{TP + FN}
\end{align}
\bigskip

Where \textit{TP} equals true positives, \textit{TN} equates to true negatives, \textit{FP} represents false positives, and finally, \textit{FN} represents the number of false negatives. \\
We also include accuracy as a supporting metric to compare classifiers, denoted by the following:    

\bigskip
\begin{align}
    accuracy = \frac{correctly\,classified\,samples}{total\,number\,of\,samples}
\end{align}
\bigskip

\subsection{Experimental results}
To validate the effectiveness of our model, we conducted two experiments: the first to evaluate the performance of the batch-incremental updating mechanism used by our system, and the second to compare the architecture of our CNN-LSTM model in an incremental learning setup with other deep learning models.

\subsubsection{Experiment 1: Comparing the performance of our technique with other incremental-learning-based ransomware detection methods}

%-------------------------------------------------------
% Figure: f2-graph
%-------------------------------------------------------
\begin{figure*}[]
    \centering
    \includegraphics[width=0.65\linewidth]{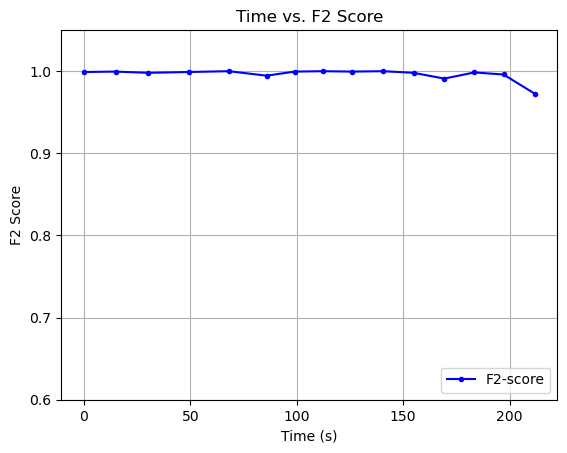}
    \caption{A graph demonstrating the F2-score vs Time of our batch-incremental ransomware detection technique (iCNN-LSTM). Each data point consists of a new training batch used to update the model.}
    \label{fig:f1-graph}
\end{figure*}
%-------------------------------------------------------

Fig. \ref{fig:f1-graph} illustrates the performance of our approach in detecting ransomware over time, measured by the F2-score. The F2-score is a weighted metric that emphasises recall over precision, making it particularly relevant in scenarios where false negatives are more damaging than false positives, such as ransomware detection.

In this graph, the F2-score consistently remains close to 1.0 throughout the entire period, with each data point corresponding to a batch update of the model. This high F2-score indicates that our approach effectively sustains a robust ability to identify new ransomware instances while accurately minimising false negatives. The fluctuations observed around the 80 and 170-second mark suggest the model encountered new ransomware samples not included in previous batches and an imbalance in the ransomware events within this segment. Despite this, the model's F2-score swiftly rebounds following an update integrating the latest data. This quick recovery highlights the model's ability to adapt and improve its classification accuracy. Identifying the optimal timing for model updates to minimise misclassification rates presents an intriguing avenue for further investigation. 

Despite continuously integrating unseen ransomware instances, the F2-score remains impressively stable, consistently exceeding 99\%. This high F2-score demonstrates the model's exceptional accuracy in classifying ransomware, indicating that the iCNN-LSTM model adapts to new data and retains a high degree of classification precision throughout the evaluation period. The consistent performance above the 99\% threshold indicates the underlying architecture's robustness and ability to generalise well in dynamic and evolving threat environments.

To assess the efficacy of our model, we compared it with other incremental learning techniques used in the literature for ransomware detection, as shown in Table \ref{tab:comparison}. However, these studies do not report either false negative rates or an F2-score. Since false negatives are more detrimental than false positives in ransomware detection, we calculated the relative F2-scores based on the recall rates, F1-scores, and precision values using Formula. \ref{eqn:f2-score}. Our approach achieved the highest F1 and F2-scores of 99.61\%.

%-------------------------------------------------------
% Table : Comparison of incremental learning techniques
%-------------------------------------------------------
\begin{table}[h!]
\centering
\caption{We compare our iCNN-LSTM model with other incremental learning techniques for ransomware detection found in the literature. We report both the F1-score and the calculated F2-score, emphasising maximising recall to prioritise the reduction of false negatives.}
\label{tab:comparison}
\begin{tabular}{|c|l|c|c|c|}
\hline
\textbf{Study}                     & \textbf{F1-score} & \textbf{F2-score} & \textbf{Recall}  & \textbf{Precision} \\ \hline
Roy and Chen \cite{roy2021deepran} & 99.39\%           & 99.03\%           & 98.8\%           & 100\%              \\ \hline
Al-rimy et al. \cite{al2019crypto} & 98.70\%           & 99.02\%           & 99.24\%          & 98.16\%            \\ \hline
\textbf{iCNN-LSTM (this study)}    & \textbf{99.61\%}  & \textbf{99.61\%}  & \textbf{99.62\%} & \textbf{99.61\%}   \\ \hline
\end{tabular}
\end{table}
%-------------------------------------------------------

\begin{table*}[htp!]
\centering
\caption{A comparison of various Cnn and LSTM ransomware detection models throughout the literature. This table presents the original results as reported in their respective studies.}
\label{tab:recorded-results}
\resizebox{0.70\linewidth}{!}{%
\begin{tabular}{|c|c|c|c|c|}
\hline
\textbf{Model}                                                                                                                                                   & \textbf{Accuracy} & \textbf{F1-score} & \textbf{Recall} & \textbf{Precision} \\ \hline
\begin{tabular}[c]{@{}c@{}}3 LSTM modules stacked sequentially with \\ 64 LSTM nodes (Maniath et al. \cite{maniath2017deep})\end{tabular}                        & 99.67\%           & n/a               & n/a             & n/a                \\ \hline
\begin{tabular}[c]{@{}c@{}}Combined single layer  CNN and LSTM model \\ with 1500 units (Agrawal et al. \cite{agrawal2018robust})\end{tabular}                   & 95.6\%            & n/a               & n/a             & n/a                \\ \hline
\begin{tabular}[c]{@{}c@{}}Single layer LSTM model with 8 units \\ (Homayoun et al. \cite{homayoun2019drthis})\end{tabular}                                      & 99.6\%            & n/a               & n/a             & n/a                \\ \hline
\begin{tabular}[c]{@{}c@{}}Multi-layer CNN modules with Attention \\ mechanisms (Zhang et al. \cite{zhang2020ransomware})\end{tabular}                           & 89.5\%            & 87.3\%            & 87.6\%          & 87.5\%             \\ \hline
\begin{tabular}[c]{@{}c@{}}Combined single layer CNN and LSTM model \\ with 512 units (Akhtar and Feng \cite{akhtar2022detection})\end{tabular}                  & 99\%              & 99\%              & 99\%            & 99\%               \\ \hline
\begin{tabular}[c]{@{}c@{}}Tri-layer CNN and LSTM modules sequentially \\ stacked (Bensaoud and Kalita \cite{bensaoud2024cnn})\end{tabular}                      & 99.91\%           & 99.62\%           & 99.62\%         & 99.62\%            \\ \hline
\begin{tabular}[c]{@{}c@{}}Multi-layer CNN, LSTM and Transpose \\ layers sequentially stacked \\ (Deivakani et al. \cite{deivakani2024intelligent})\end{tabular} & 99\%              & n/a               & 98.9\%          & 98.9\%             \\ \hline
\end{tabular}%
}
\end{table*}

\subsubsection{Experiment 2: Assessing the efficiency of our iCNN-LSTM architecture compared to other CNN and LSTM models within our batch-incremental framework}

%-------------------------------------------------------
% Figure: Study comparison
%-------------------------------------------------------
\begin{figure*}[hbp!]
    \centering
    \includegraphics[width=0.90\linewidth]{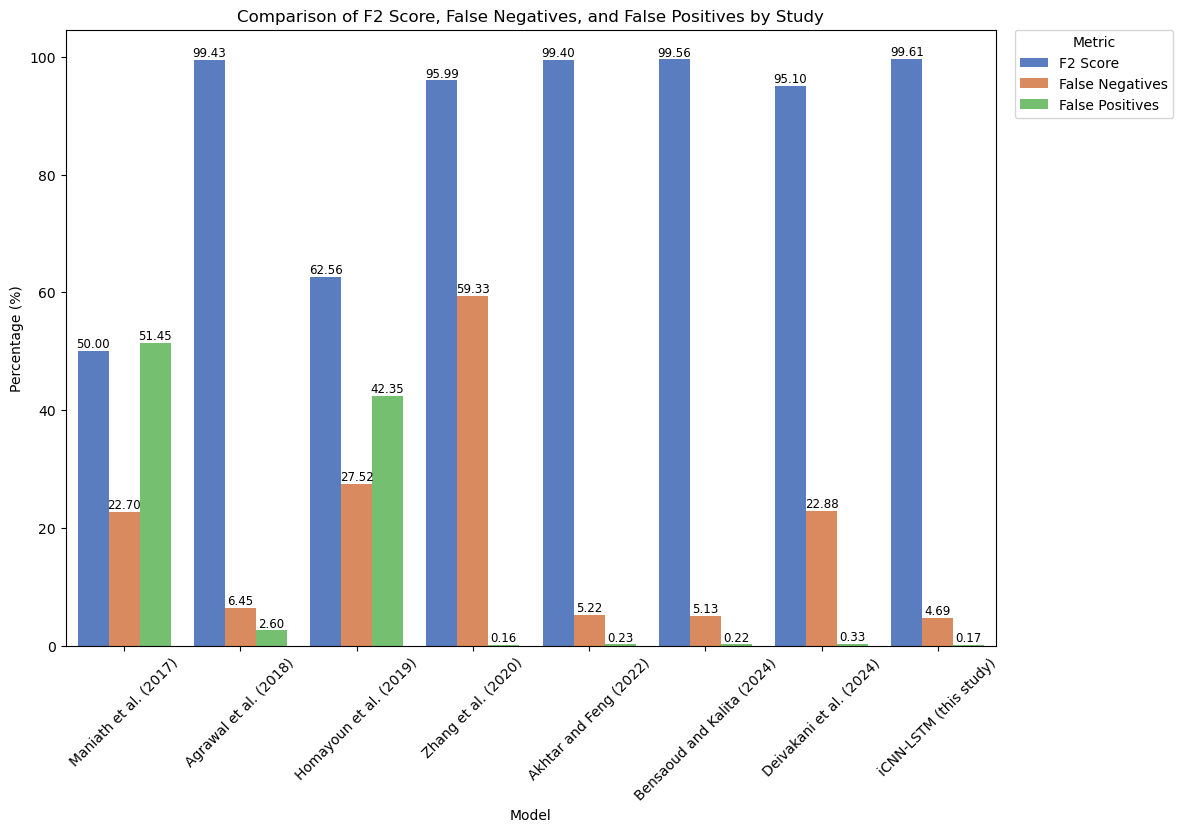}
    \caption{A comparison of the F2-score, false negatives, and false positives of various CNN/LSTM models for ransomware/malware detection, as applied within our incremental learning framework, based on the literature. }
    \label{fig:F2-graph}
\end{figure*}
%-------------------------------------------------------

%-------------------------------------------------------
% Figure: F2-score, FNR and runtime
%-------------------------------------------------------

\begin{figure*}[hbp!]
    \centering
    \includegraphics[width=0.80\linewidth]{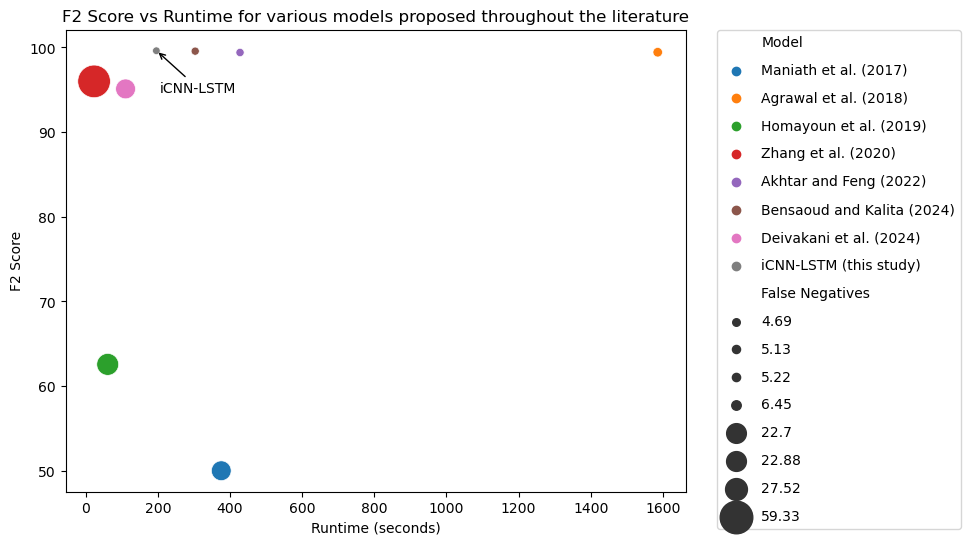}
    \caption{A comparison of the F2-score vs the runtime for various studies. The size of the data points represents the false negative rate }
    \label{fig:false-negatives}
\end{figure*}
%-------------------------------------------------------

%\--------------------------------------------
% TABLE Results
%\--------------------------------------------

\begin{table*}[htp!]
\centering
\caption{This table compares our iCNN-LSTM technique against various ransomware detection models. It presents the results of running each model from the respective studies within our batch-incremental framework on our imbalanced Sysmon dataset.}
\label{tab:experiment-results}
\resizebox{0.9\linewidth}{!}{%
\begin{tabular}{|c|c|c|c|c|c|c|c|}
\hline
\textbf{Model}                                                                                                                                                   & \textbf{F1-score} & \textbf{F2-score} & \textbf{Recall}  & \textbf{Precision} & \textbf{\begin{tabular}[c]{@{}c@{}}False\\ Positives\end{tabular}} & \textbf{\begin{tabular}[c]{@{}c@{}}False\\ Negatives\end{tabular}} & \textbf{Runtime} \\ \hline
\begin{tabular}[c]{@{}c@{}}3 LSTM modules stacked sequentially with \\ 64 LSTM nodes (Maniath et al. \cite{maniath2017deep})\end{tabular}                        & 57.23\%           & 50\%              & 53.17\%          & 61.96\%            & 51.45\%                                                            & 22.70\%                                                            & 375.67s          \\ \hline
\begin{tabular}[c]{@{}c@{}}Combined single layer  CNN and LSTM model \\ with 1500 units (Agrawal et al. \cite{agrawal2018robust})\end{tabular}                   & 99.43\%           & 99.43\%           & 99.44\%          & 99.43\%            & 2.6\%                                                              & 6.45\%                                                             & 1585.54s         \\ \hline
\begin{tabular}[c]{@{}c@{}}Single layer LSTM model with 8 units \\ (Homayoun et al. \cite{homayoun2019drthis})\end{tabular}                                      & 70.49\%           & 62.56\%           & 65.71\%          & 76.03\%            & 42.35\%                                                            & 27.52\%                                                            & 60.80s           \\ \hline
\begin{tabular}[c]{@{}c@{}}Multi-layer CNN modules with Attention \\ mechanisms (Zhang et al. \cite{zhang2020ransomware})\end{tabular}                           & 95.27\%           & 95.99\%           & 95.49\%          & 95.06\%            & 0.16\%                                                             & 59.33\%                                                            & 23.05s           \\ \hline
\begin{tabular}[c]{@{}c@{}}Combined single layer CNN and LSTM model \\ with 512 units (Akhtar and Feng \cite{akhtar2022detection})\end{tabular}                  & 99.41\%           & 99.40\%           & 99.41\%          & 99.41\%            & 0.23\%                                                             & 5.22\%                                                             & 427.62s          \\ \hline
\begin{tabular}[c]{@{}c@{}}Tri-layer CNN and LSTM modules sequentially \\ stacked (Bensaoud and Kalita \cite{bensaoud2024cnn})\end{tabular}                      & 99.56\%           & 99.56\%           & 99.56\%          & 99.56\%            & 0.22\%                                                             & 5.13\%                                                             & 303.35s          \\ \hline
\begin{tabular}[c]{@{}c@{}}Multi-layer CNN, LSTM and Transpose \\ layers sequentially stacked \\ (Deivakani et al. \cite{deivakani2024intelligent})\end{tabular} & 95.76\%           & 95.10\%           & 95.26\%          & 96.27\%            & 0.33\%                                                             & 22.88\%                                                            & 110.31s          \\ \hline
\textbf{iCNN-LSTM}                                                                                                                                               & \textbf{99.61\%}  & \textbf{99.61\%}  & \textbf{99.62\%} & \textbf{99.61\%}   & \textbf{0.17\%}                                                    & \textbf{4.69\%}                                                    & \textbf{195.69s} \\ \hline
\end{tabular}%
}
\end{table*}
%\-------------------------------------------

Several studies have utilised combined CNN-LSTM architectures throughout the literature for ransomware detection. However, we have developed a new architecture incorporating parallel LSTM modules and attention mechanisms to enhance classification speed, often limited by sequential LSTM processing. To evaluate the performance of our architecture against other techniques, we compared our iCNN-LSTM model with seven other studies that used stand-alone CNN or LSTM models or combined CNN-LSTM architectures, as shown in Table. \ref{tab:recorded-results}. As these studies did not incorporate our batch-incremental framework for incremental model updates or our dataset for evaluation, we implemented their model architectures within our framework and dataset to compare their classification performance.
As illustrated in Table. \ref{tab:experiment-results} and Fig. \ref{fig:F2-graph}, our approach consistently achieved an outstanding F2-score above 99\%, indicating a solid balance between precision and recall. This performance metric is critical in ransomware detection, where the cost of false negatives can be high. Additionally, our approach maintained the lowest false positive rate of 4.69\%, further underscoring its effectiveness in accurately identifying ransomware with minimal errors. This high level of performance was maintained even as the model was incrementally updated, highlighting its robustness in dynamic environments. Importantly, our technique successfully detected ransomware, even though it was a minority class in a highly imbalanced dataset. \\
Some studies demonstrated improved performance when using our batch-incremental framework compared to their original results, such as Agrawal et al. \cite{agrawal2018robust} and Akhtar and Feng \cite{akhtar2022detection}. This improvement is attributed to reduced model degradation from frequent updates, the effectiveness of Sysmon logs in representing ransomware activity, and the ability of CNN-LSTM modules to capture spatial and temporal patterns. However, some studies showed significant degradation when using our batch-incremental framework and imbalanced Sysmon dataset, primarily due to high false positive and false negative rates. These studies used either CNN or LSTM models individually, which led to poorer performance. In contrast, the combined CNN-LSTM architectures demonstrated superior performance within our incremental learning framework, with lower false positive and false negative rates.\\
Notably, our technique had the shortest runtime of all studies, with a false negative rate below 20\%, showcasing our CNN-LSTM architecture's efficiency. Despite the perceived complexity of combining CNN and LSTM modules, our approach outperforms models that rely solely on CNNs or LSTMs in terms of classification performance and operational efficiency. A comparison with the work of Bensaoud and Kalita \cite{bensaoud2024cnn} and Akhtar and Feng \cite{akhtar2022detection} reveals that while their methods achieve similar F2-scores when their models are applied to our batch-incremental framework, the time required for incremental training and testing is significantly longer, as seen in Fig. \ref{fig:false-negatives}. The primary reason for the improved time efficiency in our approach is the parallel configuration of LSTM modules and attention mechanisms rather than running them sequentially. By running these modules in parallel, we effectively reduce the bottlenecks typically associated with sequential temporal data processing. Additionally, integrating attention mechanisms allows the model to focus on the most relevant parts of the data, further enhancing processing speed and accuracy. The high detection accuracy, low false negative rate, and reduced latency compared to other models make our approach well-suited for detecting ransomware in real-time data streams.

\section{Conclusion and future work}

In this article, we developed a novel batch-incremental deep learning-based ransomware detection system that integrates CNNs and LSTMs to identify ransomware activity within a stream of Sysmon logs. Unlike previous approaches, our system is continuously updated with mini-batches, enabling it to accurately detect new ransomware strains as they are introduced into the dataset. We demonstrated that our technique outperforms other methods, with the iCNN-LSTM architecture significantly improving detection accuracy, reducing false negatives, and enhancing model training and testing efficiency. Our approach achieved an F2-score of 99.61\% and a false negative rate of 4.69\%, surpassing other models in performance and speed.

However, our approach has some limitations. The model trains in fixed mini-batches and lacks mechanisms to detect and address model degradation over time. Future research will focus on developing strategies to dynamically identify and respond to model degradation. Additionally, we did not address computational resource consumption in this study, which will be explored in future work. Our approach demonstrates significant utility in accurately detecting ransomware within a data stream of Sysmon logs, making it a promising solution for real-time ransomware detection.

\section{Acknowledgement}
The work has been supported by the Cyber Security Research Centre Limited, whose activities are partially funded by the Australian Government’s Cooperative Research Centres Programme.

\bibliographystyle{elsarticle-num}
\bibliography{references} % Entries are in the refs.bib file

\begin{thebibliography}{10}
\expandafter\ifx\csname url\endcsname\relax
  \def\url#1{\texttt{#1}}\fi
\expandafter\ifx\csname urlprefix\endcsname\relax\def\urlprefix{URL }\fi
\expandafter\ifx\csname href\endcsname\relax
  \def\href#1#2{#2} \def\path#1{#1}\fi

\bibitem{ispahany2021detecting}
J.~Ispahany, R.~Islam, Detecting malicious covid-19 urls using machine learning techniques (2021).

\bibitem{beerman2023review}
J.~Beerman, D.~Berent, Z.~Falter, S.~Bhunia, A review of colonial pipeline ransomware attack, in: 2023 IEEE/ACM 23rd International Symposium on Cluster, Cloud and Internet Computing Workshops (CCGridW), IEEE, 2023, pp. 8--15.

\bibitem{dossett_2021}
J.~Dossett, \href{https://www.cnet.com/personal-finance/crypto/a-timeline-of-the-biggest-ransomware-attacks/}{A timeline of the biggest ransomware attacks} (Nov 2021).
\newline\urlprefix\url{https://www.cnet.com/personal-finance/crypto/a-timeline-of-the-biggest-ransomware-attacks/}

\bibitem{ispahany2024ransomware}
J.~Ispahany, M.~R. Islam, M.~Z. Islam, M.~A. Khan, Ransomware detection using machine learning: A review, research limitations and future directions, IEEE Access (2024).

\bibitem{bello2021detecting}
I.~Bello, H.~Chiroma, U.~A. Abdullahi, A.~Y. Gital, F.~Jauro, A.~Khan, J.~O. Okesola, S.~M. Abdulhamid, Detecting ransomware attacks using intelligent algorithms: Recent development and next direction from deep learning and big data perspectives, Journal of Ambient Intelligence and Humanized Computing 12 (2021) 8699--8717.

\bibitem{fernando2020study}
D.~W. Fernando, N.~Komninos, T.~Chen, A study on the evolution of ransomware detection using machine learning and deep learning techniques, IoT 1~(2) (2020) 551--604.

\bibitem{hemalatha2021efficient}
J.~Hemalatha, S.~A. Roseline, S.~Geetha, S.~Kadry, R.~Dama{\v{s}}evi{\v{c}}ius, An efficient densenet-based deep learning model for malware detection, Entropy 23~(3) (2021) 344.

\bibitem{karbab2023swiftr}
E.~B. Karbab, M.~Debbabi, A.~Derhab, Swiftr: Cross-platform ransomware fingerprinting using hierarchical neural networks on hybrid features, Expert Systems with Applications 225 (2023) 120017.

\bibitem{zhang2020ransomware}
B.~Zhang, W.~Xiao, X.~Xiao, A.~K. Sangaiah, W.~Zhang, J.~Zhang, Ransomware classification using patch-based cnn and self-attention network on embedded n-grams of opcodes, Future Generation Computer Systems 110 (2020) 708--720.

\bibitem{zhang2021dual}
X.~Zhang, J.~Wang, S.~Zhu, Dual generative adversarial networks based unknown encryption ransomware attack detection, IEEE Access 10 (2021) 900--913.

\bibitem{gazzan2023enhanced}
M.~Gazzan, F.~T. Sheldon, An enhanced minimax loss function technique in generative adversarial network for ransomware behavior prediction, Future Internet 15~(10) (2023) 318.

\bibitem{zahoora2022ransomware}
U.~Zahoora, A.~Khan, M.~Rajarajan, S.~H. Khan, M.~Asam, T.~Jamal, Ransomware detection using deep learning based unsupervised feature extraction and a cost sensitive pareto ensemble classifier, Scientific Reports 12~(1) (2022) 15647.

\bibitem{woralert2023hard}
C.~Woralert, C.~Liu, Z.~Blasingame, Hard-lite: A lightweight hardware anomaly realtime detection framework targeting ransomware, IEEE Transactions on Circuits and Systems I: Regular Papers (2023).

\bibitem{roy2021deepran}
K.~C. Roy, Q.~Chen, Deepran: Attention-based bilstm and crf for ransomware early detection and classification, Information Systems Frontiers 23~(2) (2021) 299--315.

\bibitem{ciaramella2023explainable}
G.~Ciaramella, G.~Iadarola, F.~Martinelli, F.~Mercaldo, A.~Santone, Explainable ransomware detection with deep learning techniques, Journal of Computer Virology and Hacking Techniques (2023) 1--14.

\bibitem{luan2019research}
Y.~Luan, S.~Lin, Research on text classification based on cnn and lstm, in: 2019 IEEE international conference on artificial intelligence and computer applications (ICAICA), IEEE, 2019, pp. 352--355.

\bibitem{yang2022adaptability}
M.~Yang, J.~Wang, Adaptability of financial time series prediction based on bilstm, Procedia Computer Science 199 (2022) 18--25.

\bibitem{shaohu2024prediction}
L.~Shaohu, W.~Yuandeng, H.~Rui, Prediction of drilling plug operation parameters based on incremental learning and cnn-lstm, Geoenergy Science and Engineering 234 (2024) 212631.

\bibitem{read2012batch}
J.~Read, A.~Bifet, B.~Pfahringer, G.~Holmes, Batch-incremental versus instance-incremental learning in dynamic and evolving data, in: Advances in Intelligent Data Analysis XI: 11th International Symposium, IDA 2012, Helsinki, Finland, October 25-27, 2012. Proceedings 11, Springer, 2012, pp. 313--323.

\bibitem{gepperth2016incremental}
A.~Gepperth, B.~Hammer, Incremental learning algorithms and applications, in: European symposium on artificial neural networks (ESANN), 2016.

\bibitem{wallace2011class}
B.~C. Wallace, K.~Small, C.~E. Brodley, T.~A. Trikalinos, Class imbalance, redux, in: 2011 IEEE 11th international conference on data mining, Ieee, 2011, pp. 754--763.

\bibitem{belouadah2020active}
E.~Belouadah, A.~Popescu, U.~Aggarwal, L.~Saci, Active class incremental learning for imbalanced datasets, in: European Conference on Computer Vision, Springer, 2020, pp. 146--162.

\bibitem{al2019crypto}
B.~A.~S. Al-rimy, M.~A. Maarof, S.~Z.~M. Shaid, Crypto-ransomware early detection model using novel incremental bagging with enhanced semi-random subspace selection, Future Generation Computer Systems 101 (2019) 476--491.

\bibitem{or2019dynamic}
O.~Or-Meir, N.~Nissim, Y.~Elovici, L.~Rokach, Dynamic malware analysis in the modern era—a state of the art survey, ACM Computing Surveys (CSUR) 52~(5) (2019) 1--48.

\bibitem{thara2019auto}
D.~Thara, B.~PremaSudha, F.~Xiong, Auto-detection of epileptic seizure events using deep neural network with different feature scaling techniques, Pattern Recognition Letters 128 (2019) 544--550.

\bibitem{lusa2012evaluation}
L.~Lusa, et~al., Evaluation of smote for high-dimensional class-imbalanced microarray data, in: 2012 11th international conference on machine learning and applications, Vol.~2, IEEE, 2012, pp. 89--94.

\bibitem{bojanowski2017enriching}
P.~Bojanowski, E.~Grave, A.~Joulin, T.~Mikolov, Enriching word vectors with subword information, Transactions of the association for computational linguistics 5 (2017) 135--146.

\bibitem{venkatesh2019review}
B.~Venkatesh, J.~Anuradha, A review of feature selection and its methods, Cybernetics and information technologies 19~(1) (2019) 3--26.

\bibitem{goodfellow2016deep}
I.~Goodfellow, Y.~Bengio, A.~Courville, Deep learning, MIT press, 2016.

\bibitem{stankovic2023convolutional}
L.~Stankovi{\'c}, D.~Mandic, Convolutional neural networks demystified: A matched filtering perspective-based tutorial, IEEE Transactions on Systems, Man, and Cybernetics: Systems 53~(6) (2023) 3614--3628.

\bibitem{hochreiter1997long}
S.~Hochreiter, J.~Schmidhuber, Long short-term memory, Neural computation 9~(8) (1997) 1735--1780.

\bibitem{zheng2021understanding}
W.~Zheng, P.~Zhao, K.~Huang, G.~Chen, Understanding the property of long term memory for the lstm with attention mechanism, in: Proceedings of the 30th ACM International Conference on Information \& Knowledge Management, 2021, pp. 2708--2717.

\bibitem{chen2019exploring}
S.~Chen, L.~Ge, Exploring the attention mechanism in lstm-based hong kong stock price movement prediction, Quantitative Finance 19~(9) (2019) 1507--1515.

\bibitem{zhang2018attention}
L.~Zhang, G.~Zhu, L.~Mei, P.~Shen, S.~A.~A. Shah, M.~Bennamoun, Attention in convolutional lstm for gesture recognition, Advances in neural information processing systems 31 (2018).

\bibitem{dai2019grow}
X.~Dai, H.~Yin, N.~K. Jha, Grow and prune compact, fast, and accurate lstms, IEEE Transactions on Computers 69~(3) (2019) 441--452.

\bibitem{zhang2020deepmal}
J.~Zhang, Deepmal: A cnn-lstm model for malware detection based on dynamic semantic behaviours, in: 2020 International Conference on Computer Information and Big Data Applications (CIBDA), IEEE, 2020, pp. 313--316.

\bibitem{ozkok2022hybrid}
F.~O. Ozkok, M.~Celik, A hybrid cnn-lstm model for high resolution melting curve classification, Biomedical Signal Processing and Control 71 (2022) 103168.

\bibitem{lu2020cnn}
W.~Lu, J.~Li, Y.~Li, A.~Sun, J.~Wang, A cnn-lstm-based model to forecast stock prices, Complexity 2020~(1) (2020) 6622927.

\bibitem{akhtar2022detection}
M.~S. Akhtar, T.~Feng, Detection of malware by deep learning as cnn-lstm machine learning techniques in real time, Symmetry 14~(11) (2022) 2308.

\bibitem{srivastava2014dropout}
N.~Srivastava, G.~Hinton, A.~Krizhevsky, I.~Sutskever, R.~Salakhutdinov, Dropout: a simple way to prevent neural networks from overfitting, The journal of machine learning research 15~(1) (2014) 1929--1958.

\bibitem{hinton2012improving}
G.~E. Hinton, N.~Srivastava, A.~Krizhevsky, I.~Sutskever, R.~R. Salakhutdinov, Improving neural networks by preventing co-adaptation of feature detectors, arXiv preprint arXiv:1207.0580 (2012).

\bibitem{optuna_2019}
T.~Akiba, S.~Sano, T.~Yanase, T.~Ohta, M.~Koyama, Optuna: A next-generation hyperparameter optimization framework, in: Proceedings of the 25th {ACM} {SIGKDD} International Conference on Knowledge Discovery and Data Mining, 2019.

\bibitem{christen2023review}
P.~Christen, D.~J. Hand, N.~Kirielle, A review of the f-measure: its history, properties, criticism, and alternatives, ACM Computing Surveys 56~(3) (2023) 1--24.

\bibitem{maniath2017deep}
S.~Maniath, A.~Ashok, P.~Poornachandran, V.~Sujadevi, P.~S. AU, S.~Jan, Deep learning lstm based ransomware detection, in: 2017 Recent Developments in Control, Automation \& Power Engineering (RDCAPE), IEEE, 2017, pp. 442--446.

\bibitem{agrawal2018robust}
R.~Agrawal, J.~W. Stokes, M.~Marinescu, K.~Selvaraj, Robust neural malware detection models for emulation sequence learning, in: MILCOM 2018-2018 IEEE Military Communications Conference (MILCOM), IEEE, 2018, pp. 1--8.

\bibitem{homayoun2019drthis}
S.~Homayoun, A.~Dehghantanha, M.~Ahmadzadeh, S.~Hashemi, R.~Khayami, K.-K.~R. Choo, D.~E. Newton, Drthis: Deep ransomware threat hunting and intelligence system at the fog layer, Future Generation Computer Systems 90 (2019) 94--104.

\bibitem{bensaoud2024cnn}
A.~Bensaoud, J.~Kalita, Cnn-lstm and transfer learning models for malware classification based on opcodes and api calls, Knowledge-Based Systems 290 (2024) 111543.

\bibitem{deivakani2024intelligent}
M.~Deivakani, M.~S. Sheela, K.~Priyadarsini, Y.~Farhaoui, An intelligent security mechanism in mobile ad-hoc networks using precision probability genetic algorithms (ppga) and deep learning technique (stacked lstm), Sustainable Computing: Informatics and Systems (2024) 101021.

\end{thebibliography}

\newpage
\section{Biography Section}
\bio{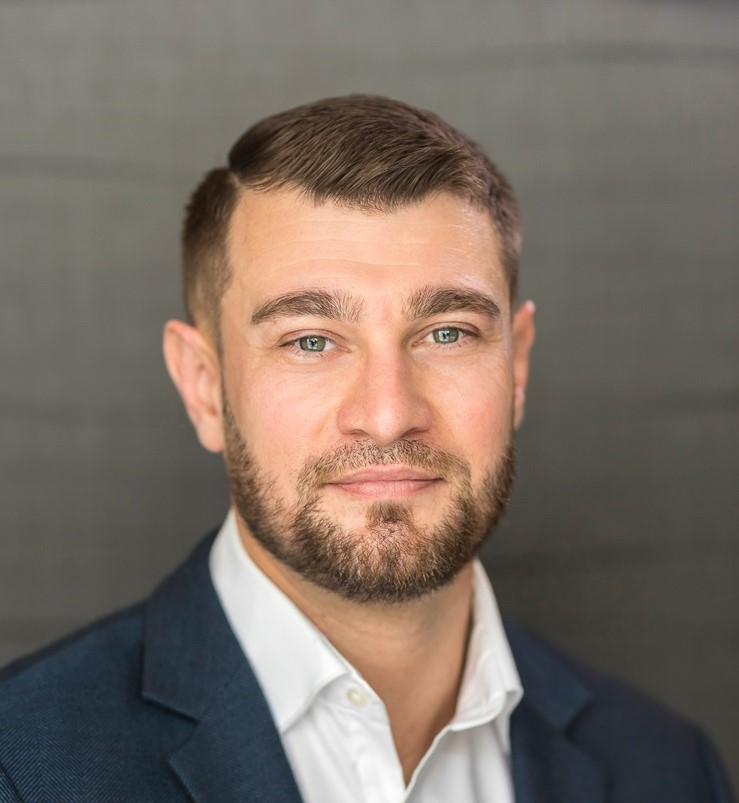}\textbf{Jamil Ispahany} is pursuing a PhD at the School of Computing, Mathematics and Engineering at Charles Sturt University, Australia. He is a recipient of a scholarship at the Cyber Security Cooperative Research Centre (CSCRC). His research interests include cyber security, machine learning and malware detection.
\\
\\
\\
\\
\\
\endbio

\bio{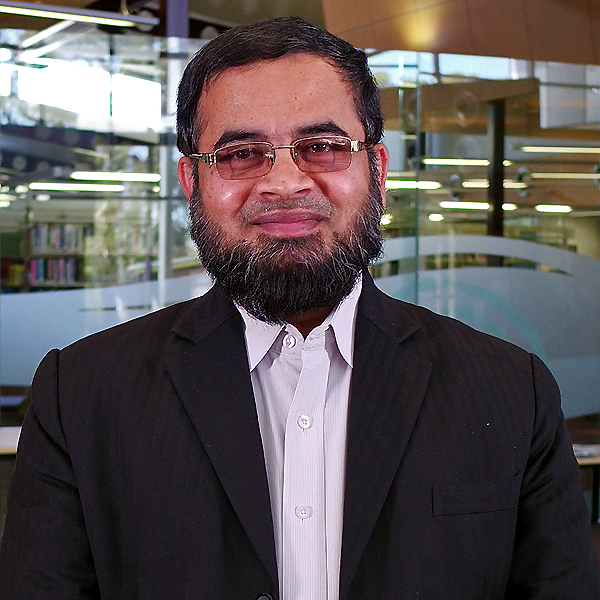}
\textbf{Md Rafiqul Islam} is working as an Associate Professor at the School of Computing, Mathematics and Engineering, Faculty of Business, Justice and Behavioral Sciences, Charles Sturt University, Australia. Dr Islam has a strong research background in Cybersecurity with a specific focus on malware analysis and classification, Authentication, security in the cloud, privacy in social media and the Internet of Things (IoT). He is leading the Cybersecurity research team and has developed a strong background in leadership, sustainability, and collaborative research. He has a strong publication record and has published more than 180 peer-reviewed research papers, book chapters and books. His contribution is recognised both nationally and internationally by achieving various rewards such as professional excellence, research excellence, and leadership awards.
\endbio

\bio{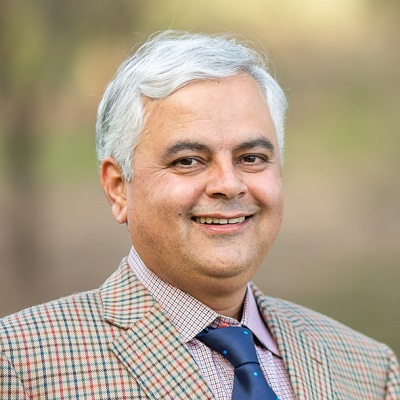}
\textbf{M. Arif Khan} received the B.Sc. degree in electrical engineering from the University of Engineering and Technology Lahore, Pakistan, the M.S. degree in electronic engineering from the GIK Institute of Engineering Sciences and Technology, Pakistan, and the Ph.D. degree in electronic engineering from Macquarie University Sydney, Australia. He is currently a Senior Lecturer with the School of Computing, Mathematics and Engineering, Charles Sturt University, Australia. His research interests include future wireless communication technologies, smart cities, massive MIMO systems, and cyber security. He was a recipient of the Prestigious International Macquarie University Research Scholarship (iMURS), and ICT CSIRO scholarships for his Ph.D. degree. He also has the competitive GIK Scholarship for his master’s degree.
\endbio

\bio{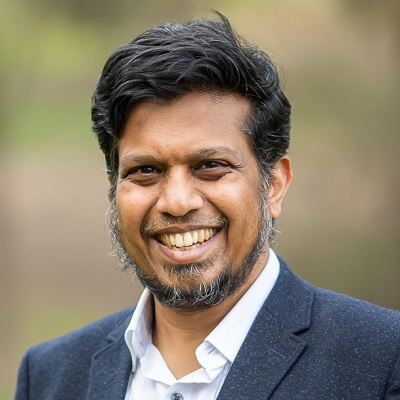}
\textbf{Md Zahidul Islam} (known as Zahid Islam) is a Professor of Computer Science, in the School of Computing, Mathematics and Engineering, Charles Sturt University, Australia. His main research interests are in Data Mining, Knowledge Discovery, Privacy Preserving Data Mining, and Applications of Data Mining/Machine Learning in various areas, including Cyber Security. URL:http://csusap.csu.edu.au/zislam/
\endbio

\end{document}